\documentclass{aa}

\usepackage{graphicx}
\usepackage{txfonts}
\usepackage{natbib}

\begin{document}

\title{SN\,2003du: 480 days in the Life of a Normal Type Ia Supernova}

\author{
V. Stanishev\inst{1}\fnmsep\thanks{E-mail: vall@physto.se}, 
A. Goobar\inst{1}, S. Benetti\inst{2}, R. Kotak\inst{3,4}, 
G. Pignata\inst{5}, H. Navasardyan\inst{2}, P. Mazzali\inst{6,7}, R. Amanullah\inst{1}, 
G. Garavini\inst{1}, 
S. Nobili\inst{1}, Y. Qiu\inst{8}, N. Elias-Rosa\inst{2,9},  P. Ruiz-Lapuente\inst{10},
J. Mendez\inst{10,11}, P. Meikle\inst{12}, F. Patat\inst{3}, A. Pastorello\inst{6,4}, 
G. Altavilla\inst{10},  M. Gustafsson\inst{13}, A. Harutyunyan\inst{2}, T. Iijima\inst{2}, 
P. Jakobsson\inst{14}, M.V. Kichizhieva\inst{15}, P. Lundqvist\inst{16}, S. Mattila\inst{12},
J. Melinder\inst{16}, E.P. Pavlenko\inst{17}, N.N. Pavlyuk\inst{18},  
J. Sollerman\inst{16,14}, D.Yu. Tsvetkov\inst{18}, M. Turatto\inst{2},
W. Hillebrandt\inst{7}
}

\institute{
Physics Department, Stockholm University, AlbaNova University Center, 106 91 Stockholm, Sweden
\and
INAF, Osservatorio Astronomico di Padova, vicolo dell'Osservatorio 5, 35122 Padova, Italy
\and
European Southern Observatory, Karl-Schwarzschild-Strasse 2, D-85748 Garching, Germany
\and
Astrophysics Research Centre, School of Mathematics and Physics,
Queen's University Belfast, BT7 1NN, UK
\and
Departamento de Astronom\'{\i}a y Astrof\'{\i}sica, Pontificia Universidad Cat\'olica
de Chile, Campus San Joaqu\'{\i}n. Vicu\~na Mackenna 4860 Casilla 306, Santiago 22, Chile
\and
INAF Osservatorio Astronomico di Trieste, Via Tiepolo 11, 34131 Trieste, Italy 
\and
Max-Planck-Institut f\"ur Astrophysik, PO Box 1317, 85741 Garching, Germany
\and
National Astronomical Observatories, Chinese Academy of Sciences, 100012 Beijing, China
\and 
Universidad de La Laguna, Av Astrof\'isico Fransisco
    S\'anchez s/n, E-38206. La Laguna, Tenerife, Spain
\and
Department of Astronomy, University of Barcelona, Marti i Franques 1, E-08028 Barcelona, Spain
\and
Isaac Newton Group of Telescopes, Apartado de correos 321, E-38700 Santa Cruz de La Palma, 
Canary Islands, Spain
\and 
Astrophysics Group, Imperial College London, Blackett Laboratory, Prince Consort Road, London, SW7 2AZ, UK
\and
Department of Physics and Astronomy, University of Aarhus, 8000 Aarhus C, Denmark
\and 
Dark Cosmology Centre, Niels Bohr Institute, University of Copenhagen,
Juliane Maries Vej 30, DK-2100 Copenhagen \O, Denmark
\and
Tavrida State University, Simferopol, Ukraine
\and
Department of Astronomy, Stockholm University, AlbaNova University Center, 106 91 Stockholm, Sweden
\and
Crimean Astrophysical Observatory, Ukraine
\and
Sternberg Astronomical Institute, Moscow State University, 
Universitetskii pr. 13, Moscow 119992,  Russia
}

\authorrunning{Stanishev et al.}

\date{Received ;accepted}

\abstract{}
{We present a study of the optical and near-infrared (NIR) properties 
of the Type Ia Supernova (SN\,Ia) 2003du.
}{
An extensive set of optical and NIR photometry and 
low-resolution long-slit spectra was obtained using a number of facilities.
The observations started 13 days before $B$-band maximum light and continued for 480 days with 
exceptionally good time sampling. The optical photometry was calibrated 
through the S-correction technique.
}{
The $UBVRIJHK$ light curves and the color indices of \object{SN\,2003du} 
closely resemble those of normal SNe\,Ia.
\object{SN\,2003du} reached a $B$-band maximum of 13.49\,$\pm$0.02 mag 
on JD2452766.38\,$\pm$0.5. We derive a $B$-band stretch 
parameter of 
% $s_B 
$0.988\,\pm$0.003, which corresponds to $\Delta 
m_{15}=1.02\,\pm0.05$, indicative of a SN\,Ia of standard luminosity. 
The reddening in the host galaxy was estimated by three methods,
and was consistently found to be negligible.
Using an updated calibration of the $V$ and $JHK$ absolute magnitudes of SNe\,Ia, 
we find a distance modulus $\mu=32.79\pm0.15$ mag to the host galaxy,
\object{UGC~9391}. We measure a peak $uvoir$  bolometric luminosity of 
$1.35(\pm0.20)\times10^{43}$\,erg\,s$^{-1}$ and  Arnett's rule implies that
 $M_{^{56}{\rm Ni}}\simeq0.68\,\pm0.14\,M_{\sun}$ of \element[][56]{Ni} was 
synthesized during the explosion. 
Modeling of the $uvoir$  bolometric light curve also indicates  
$M_{^{56}{\rm Ni}}$ in the range $0.6-0.8\,M_{\sun}$. 
The spectral evolution of \object{SN\,2003du} at both optical 
and NIR wavelengths also closely resembles normal SNe\,Ia.  In particular, 
the \ion{Si}{ii} ratio at maximum $\mathcal{R}$($\ion{Si}{ii}$)$=0.22\,\pm0.02$ and 
the time evolution of the blueshift velocities of the absorption 
line minima are typical.
The pre-maximum spectra of \object{SN\,2003du} showed conspicuous high-velocity features 
in the \ion{Ca}{ii} H\&K doublet and infrared triplet, and possibly in
\ion{Si}{ii}\,$\lambda$6355, lines. We compare 
the time evolution of the profiles of these lines with other 
well-observed SNe\,Ia and we suggest that the peculiar pre-maximum 
evolution of \ion{Si}{ii}\,$\lambda$6355 line in many SNe\,Ia is due to the presence of two
blended absorption components.
}
{}

\maketitle

\section{Introduction}
Type Ia supernovae (SNe\,Ia) form a relatively homogeneous class of objects with 
only a small scatter in their observed absolute peak magnitudes ($\sim0.3$ mag). 
Moreover, their spectra and light curves
are strikingly similar \citep[e.g.][]{bt}. 
Theoretical investigations strongly suggest that SNe\,Ia are 
thermonuclear explosions of carbon/oxygen  white dwarfs (WD) 
with masses close to the Chandrasekhar limit $\sim$\,1.4$M_\odot$
\citep[for a review see][]{hille00}. In the
favored model, the WD mass grows 
via accretion from a companion star until the mass reaches 
the Chandrasekhar limit and the WD ignites at (or near) its center. 
The light curves of SNe\,Ia are
powered by the energy released from the decay of radioactive
\element[][56]{Ni} produced during the explosion (typically a few tenths of
$M_\odot$) and its daughter nuclides, 
and the scatter of the absolute magnitudes is mostly due to
the different amounts of synthesized \element[][56]{Ni}.
However, it has been shown that the peak luminosity of SNe Ia correlates with
the luminosity decline rate after maximum light; the slower the decline, the
greater the peak luminosity \citep{psk77,dm15,ham95,ham96,riess95}. 
 After correcting for the empirical  "light curve width -- peak luminosity"
relation and for the extinction in the host galaxy,
the dispersion of the SN\,Ia absolute peak $B$ magnitudes is
$\sim0.14$ mag \citep{phil99}. This property combined with 
their high intrinsic luminosity ($M_V\simeq-19.2$ mag), make SNe\,Ia ideal for 
measuring relative cosmological distances.

Observations of SNe\,Ia  out to a redshift of $z\sim 1.0$
led to the surprising discovery that the expansion of the Universe is
accelerating, and that $\sim70$\%  of the Universe consists of an unknown constituent with 
effective negative pressure, dubbed "dark energy" 
\citep{riess98,p99,knop03,riess04,snls1,riess07,essence}.
Currently, the favored model for dark energy is a non-zero positive 
cosmological constant $\Lambda$ (or vacuum energy), but more exotic 
models  have also been proposed \citep[for a review see][]{pee}. 
There are several observational programs planned or in progress that aim to 
discover and observe  hundreds of SNe\,Ia up to $z\sim1.7$, with the goal of measuring
cosmological parameters with greatly improved accuracy. This will enable distinctions
to be made between the large number of proposed models for dark energy.  Although
these programs will be able to greatly reduce the statistical
uncertainties on the measured cosmological parameters, the output will still be limited by systematic errors
due to our poor knowledge of some aspects of SNe Ia and their
environment.
Two of the major concerns are the possible evolution of the brightness
or colors of SNe\,Ia with redshift and the estimation of the reddening
in the host galaxy.  There are indications that the amount of
\element[][56]{Ni} synthesized during the explosion is sensitive to the
metallicity, carbon-to-oxygen (C/O) ratio and the central density of
the exploding WD \citep{hof98,umeda,timmes,rop1,rop2}, although based on three-dimensional 
simulations \citet{rop1} and \citet{rop2}
found that the C/O ratio has little effect on the \element[][56]{Ni} production.
These quantities may,  however, evolve with redshift and might therefore introduce some
evolution of the observed SNe Ia properties. However, our poor knowledge of the
details of the physics of the explosion, the progenitor systems and
how the WD mass grows to the Chandrasekhar limit \citep[e.g.,][]{hille00}) 
prevents us from accurately estimating the
magnitude of the effect, and the extent to which it could affect the
derived cosmological parameters. The difficulties in accurately estimating the
reddening in the SN host galaxies arise mostly from the uncertainty in the
intrinsic colors of SNe\,Ia \citep[e.g.,][]{nob} and the calibration 
of the photometry \citep{sun_scorr}, combined with poor knowledge 
of the dust properties.

 In this paper we present observations of the nearby Type Ia  
\object{SN\,2003du}. It was discovered by The Lick Observatory  and Tenagra Observatory
Supernova Searches \citep{disc} in the nearby 
(recession velocity of 1914 km\,s$^{-1}$) SBd galaxy
\object{UGC~9391} on 2003 April 22.4 UT.
\citet{class} classified \object{SN\,2003du} as a
normal SN Ia at about two weeks before maximum light and 
an intensive optical and NIR observational campaign was initiated by the
European Supernova Collaboration (ESC). 
The optical and NIR observations were carried out until 466 and 30 days
after $B$-band maximum light, respectively; throughout this paper we
define the phase of the supernova as the time in days from the $B$-band maximum. 
The goal of the ESC is to make progress in our understanding 
of the physics of the thermonuclear  SN explosions by collecting and 
analyzing early-time observations of nearby SNe\,Ia.  
Since 2002 the ESC has obtained
via coordinated observations using a large number of telescopes 
optical and IR observations for 15 nearby SNe\,Ia. 
First results of the observations have already been published (\object{SN\,2002bo} --
\citealt{02bo}, \citealt{ab_tom};  \object{SN\,2002dj} --
\citealt{ph02dj}; \object{SN\,2002er} 
-- \citealt{ph02er}, \citealt{sp02er}; 
\object{SN\,2003cg} -- \citealt{nancy}; \object{SN\,2004eo} --
\citealt{04eo}; \object{SN\,2005cf} -- \citealt{pas05cf}, \citealt{05cf};
\citealt{ben_div,mazzali_ir}). 
Optical
observations of \object{SN\,2003du}  have also been presented by \citet{ger_du}, 
\citet{anu_du} and \citet{leo_du}.

\begin{table}[!t]
\caption{Log of the optical spectroscopy}%
\begin{tabular}{@{}lcrcc@{}}
\hline
\hline
Date (UT) & JD & Phase & Wavelength & Telescope$^a$ \\
&  &  [day] & range [\AA] & \\
\hline
2003 Apr 23 & 2452753.58  &  $-$12.8    & 4900-7500  & INT \\
2003 Apr 25 & 2452755.43  &  $-$10.9    & 3450-10400 & AS1.8  \\
2003 Apr 25 & 2452755.58  &  $-$10.8    & 4250-7500  & NOT  \\
2003 Apr 28 & 2452758.54  &  $-$7.8     & 3200-7500  & INT \\
2003 Apr 30 & 2452760.56  &  $-$5.8     & 3230-8060  & TNG \\
2003 May 02 & 2452762.39  &  $-$4.0     & 3230-8060  & TNG \\
2003 May 04 & 2452764.48  &  $-$1.9     & 3436-7776  & AS1.8  \\
2003 May 05 & 2452765.39  &  $-$1.0     & 3500-7776  & AS1.8  \\
2003 May 06 & 2452766.39  &   +0.0      & 3447-7776  & AS1.8  \\
2003 May 07 & 2452767.55  &   +1.2      & 3500-9590  & AS1.8  \\
2003 May 08 & 2452768.54  &   +2.2      & 5860-7060  & AS1.2 \\
2003 May 09 & 2452769.54  &   +3.2      & 3500-10010 & AS1.8  \\
2003 May 10 & 2452770.64  &   +4.3      & 3300-10000 & CA2.2  \\
2003 May 12 & 2452772.57  &   +6.2      & 4680-7017  & AS1.2  \\
2003 May 13 & 2452773.61  &   +7.2      & 3300-7200  & NOT \\
2003 May 14 & 2452774.59  &   +8.2      & 3300-7200  & NOT \\
2003 May 15 & 2452775.48  &   +9.1      & 3250-7200  & NOT \\
2003 May 16 & 2452776.47  &   +10.0     & 3260-9800  & NOT \\
2003 May 21 & 2452781.51  &   +15.1     & 3800-6130  & AS1.2  \\
2003 May 23 & 2452783.55  &   +17.2     & 3600-10100 & AS1.8  \\
2003 May 24 & 2452784.60  &   +18.2     & 4260-6595  & AS1.2 \\
2003 May 25 & 2452785.39  &   +19.0     & 3700-7776  & AS1.8  \\
2003 May 27 & 2452787.52  &   +21.1     & 3400-8830  & CA2.2  \\
2003 Jun 01 & 2452792.52  &   +26.1     & 3240-8060  & TNG \\
2003 Jun 06 & 2452797.60  &   +31.2     & 3240-8060  & TNG \\
2003 Jun 09 & 2452800.45  &   +34.1     & 3500-9500  & WHT \\
2003 Jun 14 & 2452805.38  &   +39.0     & 3700-10000 & WHT \\
2003 Jun 20 & 2452811.54  &   +45.2     & 3880-7770  & AS1.8  \\
2003 Jun 26 & 2452817.52  &   +51.1     & 3350-10000 & WHT \\
2003 Jul 08 & 2452829.44  &   +63.1     & 3700-9850  & NOT \\
2003 Jul 17 & 2452838.41  &   +72.0     & 3500-10000 & WHT \\
2003 Jul 29 & 2452850.42  &   +84.0     & 3600-10000 & WHT \\
2003 Aug 23 & 2452875.32  &   +108.9    & 3000-7820  & AS1.8  \\
2003 Sep 25 & 2452907.82  &   +141.4    & 3500-8800  & CA2.2 \\
2003 Nov 18 & 2452962.25  &   +195.9    & 4370-7050  & WHT$^b$ \\
2003 Dec 01 & 2452975.70  &   +209.3    & 3000-7600  & CA3.5 \\
2003 Dec 13 & 2452987.72  &   +221.3    & 3500-8820  & CA2.2 \\
2004 Feb 02 & 2453038.71  &   +272.3    & 3800-8000  & CA3.5 \\
2004 May 17 & 2453143.30  &   +376.9    & 3500-8060  & TNG \\
\hline
\end{tabular} \\
\label{t:speclog:opt}
$^a$ AS1.8 = Asiago 1.82m + AFOSC; AS1.2 = Asiago 1.22m + B\&C;
TNG = TNG 3.58m + DOLORES; NOT = NOT 2.6m + ALFOSC;
CA2.2 = Calar Alto 2.2m + CAFOS; CA3.5 = Calar Alto 3.5m + MOSCA;
WHT = WHT 4.2m + ISIS; INT = INT 2.5m + IDS   \\
$^b$ average of spectra obtained on 17 and 18 Nov. 2003; these
spectra cover the ranges 4370--5220\,\AA\ and 6200--7050\,\AA\
 with dispersion 0.23\,\AA\,pixel$^{-1}$. 
\end{table}

\section{Observations and data reduction}

\subsection{Optical spectroscopy}

The optical spectroscopy log of \object{SN\,2003du} 
is given in Table\,\ref{t:speclog:opt}.
The spectra were reduced\footnote{All data 
reduction and calibration was done in
IRAF and with our own programs written in IDL. IRAF is distributed by the
National Optical Astronomy Observatories, which are operated by the
Association of Universities for Research in Astronomy, Inc., under
cooperative agreement with the National Science Foundation.}
 following the algorithm of \citep{optext}.
The images were first bias and flat-field corrected. The 1D spectra
were then optimally extracted from the 2D images, simultaneously
identifying and removing the cosmic rays and bad pixels.
The spectra were wavelength calibrated using arc-lamp spectra.  The
wavelength calibration was checked against the night-sky emission
lines and, when necessary, small additive corrections were applied.
Spectrophotometric standard stars were used to flux calibrate the 
SN spectra. Telluric absorption features were removed from the supernova spectra 
following \citet{tell}. On a number of nights two different spectrometer
settings were used to cover the whole optical wavelength range, and
the two spectra were combined into a single
spectrum. Most of the spectra have dispersion between 
$\sim1$\,\AA\,pixel$^{-1}$ and $\sim5$\,\AA\,pixel$^{-1}$, except for the few red spectra 
taken at Asiago 1.82m telescope,  which have a dispersion of $\sim15$\,\AA\,pixel$^{-1}$ and 
one WHT spectrum with $\sim0.23$\,\AA\,pixel$^{-1}$.

The spectra were obtained with the slit oriented along the parallactic
angle in order to minimize differential losses due to atmospheric
refraction \citep{diff}. Nevertheless the relative flux calibration was not
always sufficiently accurate and
the final flux calibration was achieved by
slightly correcting the spectra to match 
the observed photometry. This step was done alongside the
calibration of the photometry and  is discussed in detail in the Appendix.

\begin{figure}[!t]
\centering
\includegraphics*[width=8.8cm]{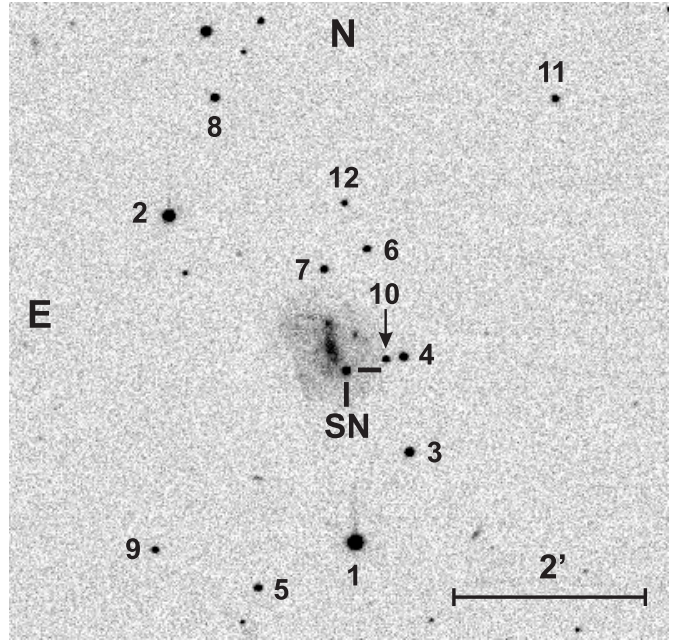}
\caption{A $B$-band finding chart of \object{SN 2003du} with the
 comparison stars labeled by numbers. The image was obtained 87 days
 after $B$-band maximum.}
\label{f:chart}
\end{figure}

\subsection{Optical photometry}

The optical photometric observations of \object{SN\,2003du} were
obtained with a number of instruments equipped with broadband $UBVRI$
filters.  The CCD images were bias and flat-field corrected.  Cosmic
ray hits were identified and cleaned with the Laplacian detection
algorithm of \citet{crs}. The observations consist of single
exposures at early times and dithered multiple exposures at late
epochs.  In the latter case, the images in each filter were 
combined to form a single image.  For the $I$-band,  we also
corrected for fringing in the individual exposures.

\begin{table}[t]
\caption{Calibrated magnitudes of the local stars around \object{SN\,2003du}. The
number in parentheses are the uncertainties in mmag.}%
\begin{tabular}{@{}r@{\hspace{1.8mm}}c@{\hspace{1.8mm}}c@{\hspace{1.8mm}}c@{\hspace{1.8mm}}c@{\hspace{1.8mm}}c@{}}
\hline
\hline
Star & $U$ & $B$ & $V$ & $R$ & $I$ \\
\hline
1  & 13.864 (39) & 13.848 (22) & 13.309 (13) & 12.960 (13) & $\cdots$	\\
2  & 15.004 (39) & 14.920 (22) & 14.310 (14) & 13.911 (13) & 13.624 (12) \\
3  & 16.562 (40) & 16.428 (22) & 15.792 (13) & 15.400 (13) & 15.077 (13) \\
4  & 16.930 (40) & 17.024 (23) & 16.462 (14) & 16.113 (14) & 15.792 (12) \\
5  & 18.261 (45) & 17.611 (24) & 16.251 (14) & 15.258 (15) & 14.117 (13) \\
6  & 18.254 (45) & 17.909 (24) & 17.011 (15) & 16.478 (14) & 16.012 (12) \\
7  & 17.660 (42) & 17.552 (23) & 16.893 (14) & 16.468 (14) & 16.129 (12) \\
8  & 17.114 (40) & 16.993 (23) & 16.307 (15) & 15.829 (14) & 15.504 (12) \\
9  & 17.806 (42) & 17.951 (24) & 17.518 (15) & 17.179 (16) & 16.875 (13) \\
10 & 17.809 (42) & 18.092 (24) & 17.675 (16) & 17.357 (16) & 17.057 (13) \\
11 & 17.775 (42) & 17.586 (23) & 16.874 (15) & 16.418 (15) & 16.107 (14) \\
12 & 18.328 (46) & 18.636 (27) & 18.158 (18) & 17.799 (18) & 17.487 (14) \\
\hline
\end{tabular}
\label{t:secstar} 
\end{table}

The SN lies only 15\arcsec\ from the host galaxy nucleus, on a 
complex background (Fig.\,\ref{f:chart}). The background 
contamination may significantly degrade the photometry, especially 
at late epochs when the SN has faded considerably. The 
approach commonly used is to subtract the background using template
galaxy images without the SN, taken either before or a few years after the SN
explosion.  The galaxy template, preferably with better seeing and
signal-to-noise ratio (S/N) than the SN images, is aligned with the
SN image, convolved with a suitable kernel so that the point-spread
functions (PSF) of the two images are the same, then scaled to match the 
flux level of the SN image and subtracted. The SN flux can then be correctly 
measured on the background-subtracted image.

Lacking pre-explosion observations of the host galaxy of 
\object{SN\,2003du}, we constructed template images using
observations which we obtained more than one year after SN maximum light.
The SN magnitudes were measured by PSF-fitting.
The small SN contribution was then subtracted
and the images were visually inspected for over- or under-subtraction
(none was noticed). The best seeing images were then combined to form
the templates. The subtraction of the host galaxy from 
the "SN + host" images was done with Alard's \citep{optsub1,optsub2} optimal image subtraction
software, slightly modified and kindly made available to us by B.
Schmidt.  
When using galaxy templates built in this way, 
any improperly subtracted SN light will introduce systematic errors into
the subsequent photometry.  In the case of \object{SN\,2003du} this
should, however, be negligible because at the epochs used to build the
templates, the SN was much fainter than on the images to which the 
template subtraction was applied (at least 2 mag fainter at $+220$ days and
4--5 mag fainter over the first three months after maximum). 
Even if we conservatively assume that the final templates still contained 20\% of the
SN light, the error introduced would be at most 0.03 mag at $+220$ days and  
clearly negligible during the first 3-4 months after
maximum.

The SN magnitudes were measured differentially with respect to 
the field stars indicated with numbers in Fig.\,\ref{f:chart}.
The instrumental magnitudes were measured by 
aperture photometry for observations before September 2003 and by PSF
fitting at the later epochs.
The magnitudes of the field stars
were calibrated for two photometric nights at the Nordic Optical Telescope -- May 15 and 16, 2003.
On each night, the field of the globular cluster \object{M92} that
includes the standard stars listed in \citet{m92} was
observed at four airmasses between 1.06 and 1.8.  The $BVRI$
magnitudes of the standard stars were taken from \citep{stetson}\footnote{Available at {\tt
http://cadcwww.hia.nrc.ca/standards/} and  as discussed by
\citet{stetson} this photometry is essentially in the  \citet{landolt92}
system.}, while the $U$ magnitudes were
calculated from the $U-B$ values given in 
\citet{m92}.  The
standard star magnitudes were measured with PSF photometry and aperture
corrections were applied to convert the PSF magnitudes to magnitudes
in an aperture with a radius of five times the seeing.  Following
\citet{harris81}, all measured magnitudes were fitted simultaneously (with 3$\sigma$
clipping) to
derive linear transformation equations, with the additional requirement that 
the color-terms and
the zero-points to be the same for the two nights.  Second-order
extinction terms were not included.  The calibrations for the two nights agree very well 
within the estimated photometric (statistical) errors.
The weighted average magnitudes  from the calibration in the two nights and 
the corresponding errors are given in Table\,\ref{t:secstar}.  
Note that the uncertainties of the calibrated magnitudes
are
donated by the uncertainty of the zero-point and not by the
statistical uncertainty.
A comparison between the stars in common with \citet{leo_du}  and \citet{anu_du} 
reveals that there are small systematic differences between the photometry; ours
being generally brighter. The mean differences with the $BVRI$ photometry of
\citet{leo_du} are, respectively, $0.010\pm0.020$, $0.013\pm0.020$, $0.039\pm0.010$ and
$0.013\pm0.027$ mag. Excluding star \#1 which is brighter in \citet{anu_du} 
in all bands, the mean differences are $0.00\pm0.05$, $0.06\pm0.01$, $0.04\pm0.01$, 
$0.06\pm0.02$ and $0.015\pm0.010$ mag for the $UBVRI$ bands, respectively. 
Some of these differences are non-negligible and we have no explanation of why they 
appear in the comparison stars calibrations. This is clearly worrisome and
emphasizes one important source of systematic errors when different SN data sets
are combined and used to derive cosmological parameters.

\citet{landolt92} standard fields were observed to derive
the instrument color-terms ($ct$), allowing us to transform the photometry
of \object{SN\,2003du} to the standard Johnson-Cousins system. The
instrumental magnitudes of the standard stars were measured by aperture
photometry with large apertures. All measurements for a given instrument 
were fitted simultaneously (with 3$\sigma$
clipping) with linear equations of the form:
\begin{eqnarray}
 U-u=ct_U(U-B)+zp & , & B-b=ct_B(B-V)+zp \nonumber \label{eq:tran} \\
V-v=ct_V(B-V)+zp & , & R-r=ct_R(V-R)+zp \\
I-i=ct_I(V-I)+zp         &   &  \nonumber
%\label{eq:tran} 
\end{eqnarray}
to determine the $ct$s.
The upper-case and lower-case letters denote the standard and 
instrumental magnitudes, respectively.

For each SN image, a zero-point was calculated for each calibrated star
 by applying Eqs.\,\ref{eq:tran}. 
The final image ZP and its uncertainty are, respectively, the average
of the individual ZPs (with $3\sigma$ outliers removed if present) and the
standard deviation. The measured scatter for the brightest stars was always 
larger than expected from Poisson statistics.
This indicates that there are additional sources of uncertainties:
imperfect flat-fielding,
presence of non-uniform scattered light, CCD non-linearity, etc.
Considering the magnitude scatter of the brightest stars we estimate that these
effects contribute $\leq0.01$ mag to the error budget.  Finally, the
ZPs were added to the measured SN magnitudes to obtain the magnitudes
in the natural systems of the instruments used, $m^{nat}$.

 \begin{figure*}[!t]
 \sidecaption
 \includegraphics*[width=12cm]{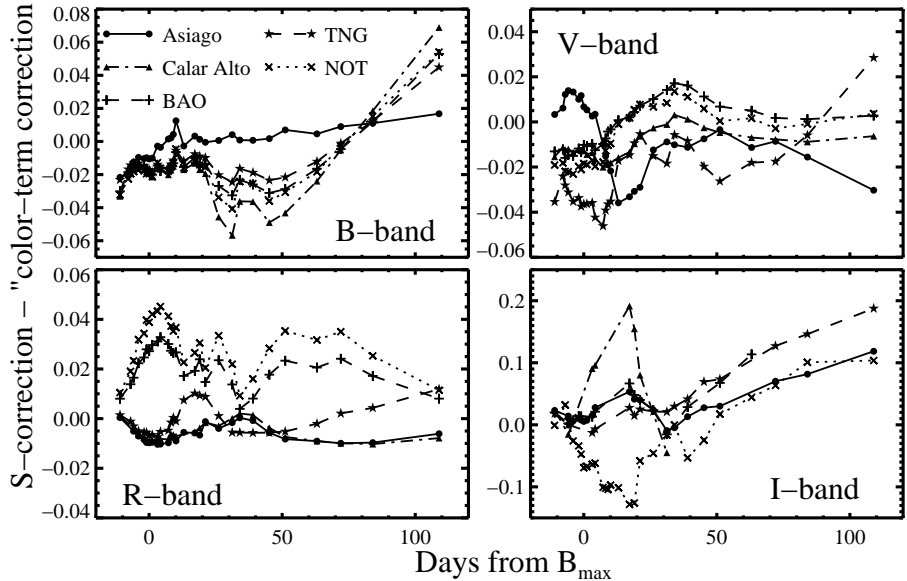}
 \caption{Time evolution of the difference between the S-correction 
 and the linear color-term correction.}
 \label{f:scorrs}
 \end{figure*}

The SN magnitudes can be  transformed to a standard photometric system
using the color
corrections obtained with Eqs.\,\ref{eq:tran}.  It is, however,
well known that these color corrections do not work well for SNe and
significant systematic differences between photometry obtained with
different instruments are often observed \citep{sun_scorr,phot99ee,kri_01el}. 
 The reason is that the SN spectral energy
distribution (SED) is very different from that of normal
stars.  Another consequence of this is that if a given band is
calibrated against different color indices, e.g. $V(B-V)$ and
$V(V-R)$, one would get the same magnitude for normal stars but
slightly different magnitudes for objects with non-stellar SEDs. This
is because the color-terms are determined with normal stars, but SNe occupy a
different region in the color-color diagrams.
The photometric observations of \object{SN\,2003du} were collected with many 
different instruments and we chose to standardize the photometry 
using the S-correction method described by \citet{phot99ee}
coupled with our very well-sampled spectral sequence of \object{SN\,2003du}.
 The S-correction method assumes that the SED of the SN and the
response of the instruments used for the observations are both
accurately known. Then one can correct the photometry to any well-defined
photometric system by means of synthetic photometry.
If $f_\lambda^{phot}(\lambda)$ is the photon flux of the object per unit
wavelength, $m^{nat}$ the object magnitude as defined  above, 
$R^{nat}(\lambda)$ the response of the natural system
and $R^{std}(\lambda)$ the response of the standard system, then the
object standard magnitude $m^{std}$ is:
\begin{eqnarray}
m^{std}=m^{nat} & -2.5\log\left(\int\,f_\lambda^{phot}(\lambda)R^{std}(\lambda)d\lambda\right) & \nonumber\\
 & +2.5\log\left(\int\,f_\lambda^{phot}(\lambda)R^{nat}(\lambda)d\lambda\right) & + const
\label{eq:scorr}
\end{eqnarray}
The constant in Eq.\,\ref{eq:scorr} is such that the correction is
zero for A0\,V stars with all color indices zero. This ensures that
for normal stars the synthetic S-correction gives the same results as
the linear color-term corrections (Eq.\,\ref{eq:tran}). The constant
can be determined from synthetic photometry of stars for which both
photometry and spectrophotometry is available.
 The details of the application of the S-corrections are given in the 
Appendix. In Fig.\,\ref{f:scorrs} we only show  the time evolution of the difference
between the S-correction and the linear color-term correction.  
Note the particularly large difference for Calar Alto
$I$, and NOT $R$ and $I$-bands, as well as the rather large scatter
for the $V$-band at all epochs and for the $B$-band  after +20 days.

\begin{figure}[!t]
 \centering
 \includegraphics*[width=8cm]{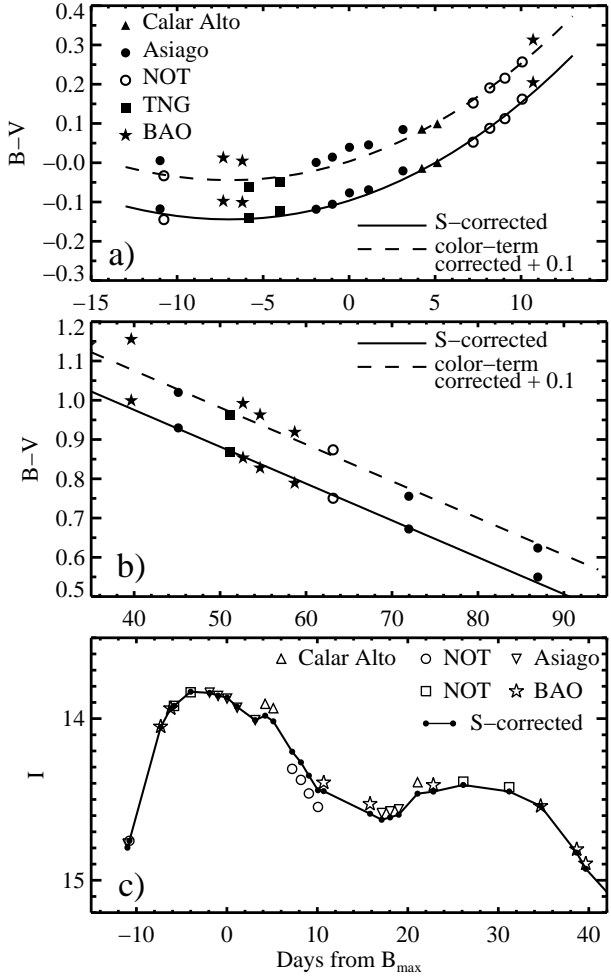}
 \caption{Comparison between the S-corrected and color-term corrected 
 {\bf a-b} $B-V$ color index and {\bf c} $I$-band magnitude. The color-term corrected
 $B-V$ data are shifted by 0.1 mag.  A polynomial fit to the S-corrected  
 $B-V$ data is overplotted. To highlight the differences the fit is also plotted  
 shifted by 0.1 mag.}
 \label{f:scor_comp}
 \end{figure}

 The final photometry of \object{SN\,2003du} is given in Table\,\ref{t:lcopt}. 
Note that none of the $U$-band and part of the $BVRI$ photometry could be 
S-corrected. Additional $B$ and $V$ photometry obtained at Moscow and Crimean
Observatories is given in Table\,\ref{t:lcopt_add}. 
Figure\,\ref{f:scor_comp} shows a comparison between the
S-corrected and color-term corrected $B-V$ color index and $I$
magnitudes. 
It is evident that in the color-term corrected photometry small systematic
differences between the various setups exist. It is also evident that the
S-correction removes those differences to a large extent, the 
exception being the BAO data at early epochs.
Significant improvement is also achieved for the $I$-band, which
required the largest S-corrections.

\begin{table*}
\caption{Optical photometry \object{SN\,2003du}.
The measurements on the dates marked with "$\ast$" and all $U$ magnitudes are not
S-corrected.}%
\begin{tabular}{@{}lc@{\hspace{3mm}}c@{\hspace{3mm}}c@{\hspace{3mm}}c@{\hspace{3mm}}c@{\hspace{3mm}}c@{\hspace{3mm}}c@{\hspace{3mm}}c@{}}
\hline
\hline
   Date (UT)   &  Phase [day] &  JD  &	 $U$   	 &  $B$	  &   $V$      &     $R$    & $I$		&  Telescope  \\
\hline
 2003-04-25   &  $-$11.0 & 2452755.39 &   $\cdots$	    &  14.737  (0.018)  &  14.854  (0.014)  &  14.728  (0.010)  &  14.798  (0.017)  &	AS1.8  \\
 2003-04-25   &  $-$10.8 & 2452755.61 &    14.382   (0.011)  &  14.629  (0.010)  &  14.774  (0.010)  &  14.672  (0.010)  &  14.756  (0.021)  &	NOT    \\
 2003-04-29   &  $-$7.3	 & 2452759.06 &  $\cdots$	    &  13.974  (0.014) & 14.072  (0.022) & 13.963  (0.015) & 14.052  (0.017)  &   BAO    \\
 2003-04-30   &  $-$6.2	 & 2452760.17 &  $\cdots$	    &  13.820  (0.024) & 13.920  (0.012) & 13.852  (0.010) & 13.936  (0.024)  &   BAO    \\
 2003-04-30   &  $-$5.8	 & 2452760.54 &    13.193   (0.034)  &  13.719  (0.018)  &  13.860  (0.010)  &  13.754  (0.010)  &  13.921  (0.021)  &	TNG    \\
 2003-05-02   &  $-$4.0	 & 2452762.38 &    13.077   (0.015)  &  13.589  (0.021)  &  13.712  (0.027)  &  13.639  (0.022)  &  13.834  (0.023)  &	TNG    \\
 2003-05-04   &  $-$1.9	 & 2452764.46 &   $\cdots$	    &  13.496  (0.010)  &  13.614  (0.018)  &  13.592  (0.010)  &  13.841  (0.010)  &	AS1.8  \\
 2003-05-05   &  $-$1.0	 & 2452765.41 &   $\cdots$	    &  13.489  (0.012)  &  13.595  (0.019)  &  13.569  (0.010)  &  13.857  (0.011)  &	AS1.8  \\
 2003-05-06   &   +0.0   & 2452766.40 &   $\cdots$	    &  13.489  (0.019)  &  13.566  (0.025)  &  13.575  (0.011)  &  13.870  (0.010)  &	AS1.8  \\
 2003-05-07   &   +1.1   & 2452767.51 &   $\cdots$	    &  13.506  (0.015)  &  13.575  (0.014)  &  13.590  (0.018)  &  13.927  (0.016)  &	AS1.8  \\
 2003-05-09   &   +3.1   & 2452769.51 &   $\cdots$	    &  13.566  (0.010)  &  13.587  (0.021)  &  13.600  (0.010)  &  14.009  (0.010)  &	AS1.8  \\
 2003-05-10   &   +4.2   & 2452770.61 &    13.234   (0.014)  &  13.605  (0.010)  &  13.620  (0.016)  &  13.600  (0.010)  &  13.982  (0.016)  &	CA2.2	 \\
 2003-05-11   &   +5.1   & 2452771.51 &    13.316   (0.040)  &  13.643  (0.011)  &  13.642  (0.017)  & $\cdots$  	&  14.017  (0.010)  &	CA2.2	\\
 2003-05-13   &   +7.2   & 2452773.59 &    13.597   (0.021)  &  13.764  (0.010)  &  13.712  (0.010)  &  13.786  (0.011)  &  14.205  (0.018)  &	NOT    \\
 2003-05-14   &   +8.2   & 2452774.56 &    13.651   (0.017)  &  13.845  (0.011)  &  13.758  (0.010)  &  13.838  (0.010)  &  14.271  (0.010)  &	NOT    \\
 2003-05-15   &   +9.1   & 2452775.44 &    13.749   (0.017)  &  13.908  (0.010)  &  13.795  (0.011)  &  13.908  (0.011)  &  14.352  (0.015)  &	NOT    \\
 2003-05-16   &   +10.1  & 2452776.45 &    13.838   (0.021)  &  14.004  (0.014)  &  13.842  (0.010)  &  13.979  (0.010)  &  14.443  (0.011)  &	NOT    \\
 2003-05-17   &   +10.7  & 2452777.08 &  $\cdots$	    &  14.066  (0.018) & 13.862  (0.010) & 14.040  (0.011) & 14.448  (0.010)  &   BAO    \\
 2003-05-22   &   +15.8  & 2452782.20 &  $\cdots$	    &  14.596  (0.013) & 14.186  (0.010) & 14.342  (0.011) & 14.588  (0.017)  &   BAO    \\
 2003-05-23   &   +17.1  & 2452783.49 &   $\cdots$	    &  14.722  (0.023)  &  14.268  (0.028)  &  14.332  (0.012)  &  14.626  (0.031)  &	AS1.8  \\
 2003-05-24   &   +18.0  & 2452784.42 &   $\cdots$	    &  14.833  (0.014)  &  14.302  (0.014)  &  14.350  (0.016)  &  14.611  (0.010)  &	AS1.8  \\
 2003-05-25   &   +19.0  & 2452785.37 &   $\cdots$	    &  14.942  (0.012)  &  14.350  (0.013)  &  14.361  (0.012)  &  14.594  (0.010)  &	AS1.8  \\
 2003-05-26   &   +20.0  & 2452786.38 &   $\cdots$	    &  15.043  (0.011)  & $\cdots$	    & $\cdots$  	& $\cdots$	    &	CA2.2	 \\
 2003-05-27   &   +21.0  & 2452787.45 &    15.144   (0.017)  &  15.146  (0.014)  &  14.463  (0.017)  &  14.392  (0.012)  &  14.464  (0.018)  &	CA2.2	 \\
 2003-05-29   &   +22.8  & 2452789.20 &  $\cdots$	    &	$\cdots$      &   $\cdots$	& 14.493  (0.031) & 14.451  (0.014)  &  BAO	 \\
 2003-06-01   &   +26.1  & 2452792.51 &    15.809   (0.023)  &  15.648  (0.010)  &  14.724  (0.010)  &  14.453  (0.019)  &  14.410  (0.027)  &	TNG    \\
 2003-06-06   &   +31.2  & 2452797.58 &    16.172   (0.033)  &  16.041  (0.011)  &  15.018  (0.014)  &  14.658  (0.010)  &  14.451  (0.019)  &	TNG    \\
 2003-06-10   &   +34.7  & 2452801.06 &  $\cdots$	    & 16.363  (0.085) & 15.223  (0.047) & 14.881  (0.018) & 14.540  (0.035) &	BAO	 \\
 2003-06-14   &   +38.7  & 2452805.04 &  $\cdots$	    &	$\cdots$      & 15.477  (0.131) & 15.140  (0.013) & 14.829  (0.024) &	BAO	 \\
 2003-06-15   &   +39.7  & 2452806.04 &  $\cdots$	    & 16.472  (0.036) & 15.473  (0.011) & 15.187  (0.018) & 14.929  (0.018) &	BAO	 \\
 2003-06-20   &   +45.1  & 2452811.51 &   $\cdots$	    &  16.606  (0.021)  &  15.676  (0.010)  &  15.415  (0.017)  &  15.265  (0.010)  &	AS1.8  \\
 2003-06-26   &   +51.1  & 2452817.52 &    16.796   (0.023)  &  16.727  (0.011)  &  15.859  (0.015)  &  15.613  (0.011)  &  15.584  (0.022)  &	TNG    \\
 2003-06-28   &   +52.7  & 2452819.04 &  $\cdots$	    &  16.745  (0.023) & 15.892  (0.012) & 15.679  (0.018) & 15.633   (0.038)  &   BAO    \\
 2003-06-30   &   +54.6  & 2452821.03 &  $\cdots$	    &  16.753  (0.039) & 15.924  (0.018) & 15.776  (0.032) & 15.715   (0.039)  &   BAO    \\
 2003-07-04   &   +58.7  & 2452825.06 &  $\cdots$	    &  16.835  (0.016) & 16.046  (0.019) & 15.874  (0.011) & 15.920   (0.022)  &   BAO    \\
 2003-07-05   &   +60.0  & 2452826.38 &   $\cdots$	    & $\cdots$  	&  16.088  (0.016)  &  15.926  (0.010)  &  16.026  (0.019)  &	NOT	\\
 2003-07-08   &   +62.7  & 2452829.06 &  $\cdots$	    &	$\cdots$      & 16.159  (0.036) & 16.010  (0.017) & 16.096  (0.029) &	BAO	 \\
 2003-07-08   &   +63.1  & 2452829.53 &    17.004   (0.026)  &  16.924  (0.011)  &  16.173  (0.010)  &  16.036  (0.011)  &  16.138  (0.017)  &	NOT    \\
 2003-07-09   &   +63.7  & 2452830.06 &  $\cdots$	    &	$\cdots$      &   $\cdots$	&   $\cdots$	  & 16.194  (0.045)  &  BAO	 \\
 2003-07-12$\ast$   &   +66.8  & 2452833.19 &  $\cdots$	    &  $\cdots$ 	&  16.290  (0.020)  &  16.140  (0.020)  &  16.270  (0.030)  &	MDK  \\
 2003-07-17   &   +72.0  & 2452838.33 &   $\cdots$	    &  17.082  (0.015)  &  16.410  (0.015)  &  16.275  (0.010)  &  16.477  (0.012)  &	AS1.8  \\
 2003-08-01   &   +87.0  & 2452853.34 &   $\cdots$	    &  17.319  (0.010)  &  16.769  (0.011)  &  16.719  (0.011)  &  17.010  (0.011)  &	AS1.8  \\
 2003-08-22   &   +108.0 & 2452874.40 &   $\cdots$	    &  17.618  (0.030)  &  17.266  (0.026)  &  17.309  (0.028)  & $\cdots$	    &	AS1.8	 \\
 2003-08-23   &   +109.0 & 2452875.37 &   $\cdots$	    &  17.688  (0.011)  &  17.281  (0.012)  &  17.379  (0.014)  &  17.756  (0.018)  &	AS1.8  \\
 2003-09-16$\ast$   &   +132.9 & 2452899.32 &    18.763   (0.079)  &  17.961  (0.022)  &  17.780  (0.021)  &  18.040  (0.023)  &  18.454  (0.075)  &	CA2.2	 \\
 2003-09-19$\ast$   &   +136.0 & 2452902.38 &    18.844   (0.032)  &  18.059  (0.012)  &  17.851  (0.017)  &  18.077  (0.029)  &  18.329  (0.020)  &	WHT    \\
 2003-09-26$\ast$   &   +143.0 & 2452909.34 &    19.140   (0.072)  &  18.114  (0.021)  &  17.956  (0.043)  &  18.330  (0.062)  &  18.570  (0.067)  &	CA2.2	 \\
 2003-11-22$\ast$   &   +199.2 & 2452965.62 &  $\cdots$	    &  18.660  (0.100)  &  18.670  (0.100)  &  19.090  (0.120)  &  $\cdots$	    &	CRM	 \\
 2003-11-23$\ast$   &   +200.3 & 2452966.63 &  $\cdots$	    &  18.990  (0.060)  &  18.930  (0.060)  &  19.210  (0.110)  &  $\cdots$	    &	CRM	 \\
 2003-11-25$\ast$   &   +202.2 & 2452968.61 &  $\cdots$	    &  18.900  (0.040)  &  18.950  (0.050)  &  19.330  (0.110)  &  $\cdots$	    &	CRM	 \\
 2003-12-01$\ast$   &   +208.3 & 2452974.63 &  $\cdots$	    &  19.020  (0.070)  &  19.090  (0.100)  &  19.370  (0.120)  &  $\cdots$	    &	CRM	 \\
 2003-12-12$\ast$   &   +220.4 & 2452986.75 &    20.688   (0.038)  &  19.304  (0.016)  &  19.313  (0.012)  &  19.899  (0.041)  &  19.673  (0.031)  &  CA3.5   \\
 2003-12-19$\ast$   &   +227.4 & 2452993.75 &   $\cdots$	    &  19.384  (0.010)  &  19.419  (0.011)  &  19.979  (0.044)  &  19.910  (0.033)  &  CA3.5	\\
 2004-05-10$\ast$   &   +370.2 & 2453136.62 &   $\cdots$	    &  21.476  (0.019)  &  21.453  (0.020)  &  22.202  (0.044)  &  21.192  (0.029)  &	WHT	    \\
 2004-05-11$\ast$   &   +371.2 & 2453137.62 &    22.638   (0.079)  &  21.531  (0.011)  &  21.493  (0.018)  &  22.190  (0.028)  &  21.320  (0.030)  &	WHT    \\
 2004-06-22$\ast$   &   +413.4 & 2453179.54 &   $\cdots$	    &  22.100  (0.044)  &  22.010  (0.026)  &  23.010  (0.052)  &  21.720  (0.038)  &	NOT	    \\
 2004-08-11$\ast$   &   +463.0 & 2453229.43 &   $\cdots$	   &  22.771  (0.027)  &  22.827  (0.026)  & $\cdots$	       & $\cdots$	   &   NOT     \\
 2004-08-14$\ast$   &   +466.0 & 2453232.39 &   $\cdots$	   & $\cdots$	       & $\cdots$	   & $\cdots$	       &  22.212  (0.048)  &   TNG     \\
\hline
\end{tabular}\\
AS1.8 -- Asiago 1.82m + AFOSC; NOT -- Nordic Optical Telescope + ALFOSC; CA2.2 -- Calar Alto 2.2m + CAFOS; TNG -- Telescopio Nazionale Galileo +
DOLORES; WHT -- William Herschel Telescope + PFIP; CA3.5 -- Calar Alto 3.5m + LAICA; BAO -- Beijing Astronomical Observatory 60cm + CCD; MDK -- Maidanak Observatory
1.5m + SITe CCD; CRM -- 60-cm Crimean reflector + CCD.
\label{t:lcopt}
\end{table*}

\subsection{Near infrared photometry and spectroscopy}

Near infrared $JHK$ photometry of \object{SN\,2003du} was obtained on
six nights at TNG and NOT. 
The two telescopes use identical $J$ and $H$ filters, 
but the TNG uses a $K'$ while the NOT  has a $K_s$ filter 
\citep{irfilt}.
The observations were reduced in the standard
way, using the XDIMSUM package in IRAF. 

The two nights at the NOT were photometric and standard stars from the
list of \citet{irstan} were observed in order to
calibrate a local sequence of stars. However, only star \#3
(Fig.\,\ref{f:chart}) could be reliably calibrated because it is the only one
that is faint enough to be in the linear range of the detector and
is bright enough to give an adequate S/N. 
The average NIR magnitudes of star \#3 are $J=14.67$, $H=14.38$ and
$K=14.37$, all with uncertainties of $\sim0.03$ mag.
The calibrated magnitudes are in good agreement with the 2MASS values,
which are $J=14.633\,\pm0.037$, $H=14.362\,\pm0.056$ and
$K=14.311\,\pm0.062$.  Star \#3 was used to calibrate the TNG
photometry. No color terms were applied. The NIR photometry 
of \object{SN\,2003du} is given in Table\,\ref{t:lcir}.

Eleven low-resolution NIR spectra of SN 2003du were obtained at
UKIRT and TNG (Table\,\ref{t:speclog:nir}). At UKIRT, the spectral range was covered by
using different instrument settings. At TNG an AMICI prism was used as
disperser. In this mode the whole NIR spectral range is provided in
one exposure at the expense of having very low resolving power
($\leq100$). Both sets of observations were
performed in ABBA sequences, where A and B denote two different
positions along the slit. 
After bias/dark and flat field corrections, 
for each pair of AB images, the B image was subtracted from the A
image.
The negative spectrum was shifted to the position of the positive one
and subtracted from it. This resulted in an image with the sum of the
spectra but minus the sky background. All such images were summed into
a single image and
the 1D spectra were then optimally extracted.  We note that the optimal 
extraction algorithm has to be applied on images where the pixel 
levels are given in the form of actual detected counts, and so it will
not work quite correctly if applied to background-subtracted
images. Special care was thus taken to calculate the optimal extraction
weights correctly.
The UKIRT spectra were wavelength calibrated with arc-lamp
spectra, while for the TNG spectra a tabulated dispersion solution
relating pixel number to wavelength was used. The dispersion of the UKIRT 
spectra ranges from $\sim5$\,\AA\,pixel$^{-1}$ to $\sim25$\,\AA\,pixel$^{-1}$, 
while for the TNG spectra, the dispersion is in the
$\sim30$\,\AA\,pixel$^{-1}$ -- $\sim100$\,\AA\,pixel$^{-1}$ range.

The A5\,V star AS-24 \citep{irstan} and 
the F7\,V star BS5581 (from the list of UKIRT standard stars) were observed
at TNG and UKIRT respectively. The standard stars
were observed close in time and airmass to the SN observations.
The SN spectra were first divided by the spectra of the comparison stars to 
remove the strong telluric absorption features.
The result was multiplied by a model spectrum of the 
appropriate spectral type, smoothed to the instrumental resolution, to remove
any residual features due to the absorption lines of the standard, 
simultaneously providing the relative flux calibration.
The UKIRT spectra from the different instrument settings that did not overlap 
were combined using 
the \object{SN\,2003du} photometry and average NIR color indices of
normal SNe\,Ia.

\section{Results}

\subsection{Light curves}

\begin{table}
\caption{Additional photometry \object{SN\,2003du}.}%
\begin{tabular}{@{}crccccc@{}}
\hline
\hline
  JD	    &     Phase     & $B$	 &   $V$	     &  Telescope \\
	       \hline
 2452765.38    &  $-$1.0  &   13.45 (0.01)  &  13.61 (0.01)   &   1 \\
 2452768.33    &   +2.0   &   13.49 (0.02)  &  13.57 (0.02)   &   1 \\
 2452775.38    &   +9.0   &   13.95 (0.01)  &  13.83 (0.01)   &   1 \\
 2452782.37    &  +16.0   &   14.65 (0.05)  &  14.17 (0.02)   &     2  \\
 2452786.33    &  +20.0   &   14.99 (0.06)  &  14.43 (0.02)   &     2  \\
 2452792.31    &  +25.9   &   $\cdots$      &  14.73 (0.06)   &     3  \\
\hline
\end{tabular}\\
1 -- 70-cm Moscow reflector + CCD Pictor 416; 2 -- 30-cm Moscow refractor + CCD AP-7p; 
3 -- 38-cm Crimean reflector + CCD ST-7;

\label{t:lcopt_add}
\end{table}

\begin{table}
\caption{NIR photometry of SN 2003du. The observations on 10.5 and 11.5 
days are from NOT. The other four are from TNG.}%
\begin{tabular}{@{}crccc@{}}
\hline
\hline
   JD & Phase & $J$ & $H$ & $K$  \\
    & [day] & & & \\
\hline
  2452755.41   & $-$11.5 &    14.96 (0.04) & 15.02 (0.04)   &  15.02 (0.04)  \\
  2452768.68   &  +1.7   &    14.42 (0.04) & 14.66 (0.04)   &  14.38 (0.04)  \\
  2452773.43   &  +6.5   &    14.92 (0.04) & 14.77 (0.04)   &  14.53 (0.04)  \\
  2452777.46   &  +10.5  &    15.67 (0.04) & 14.86 (0.04)   &  14.70 (0.04)  \\
  2452778.51   &  +11.5  &    15.84 (0.04) & 14.86 (0.04)   &  14.75 (0.04)  \\
  2452782.58   &  +15.6  &    16.12 (0.04) & 14.85 (0.04)   &  14.65 (0.04)  \\
\hline		
\end{tabular}	
\label{t:lcir}
\end{table}

\begin{table}
\caption{Log of the NIR spectroscopy}%
\begin{tabular}{@{}lcrcc@{}}
\hline
\hline
Date (UT) & JD & Phase & Coverage & Telescope$^a$ \\
   &   & [day] & [$\mu$m]& \\
\hline
2003 Apr 25 & 2452754.89  &  $-$11.5     &  0.8-2.5   & UK-1 \\
2003 Apr 25 & 2452755.47  &  $-$10.9     & 0.75-2.45  & TNG \\
2003 May 01 & 2452760.89  &  $-$5.5      &  0.8-2.5   & UK-1 \\
2003 May 04 & 2452763.88  &  $-$2.5      &  1.42-2.4  & UK-2 \\
2003 May 08 & 2452768.68  &   +2.3       &  0.9-2.3   & TNG \\
2003 May 10 & 2452769.79  &   +3.4       & 1.39-2.50  & UK-2 \\
2003 May 11 & 2452770.90  &   +4.5       &  0.8-2.5   & UK-1 \\
2003 May 19 & 2452778.80  &   +12.4      & 1.48-2.30  & UK-2 \\
2003 May 22 & 2452782.58  &   +16.2      &  0.9-2.48  & TNG \\
2003 May 27 & 2452786.77  &   +20.4      &  0.8-2.5   & UK-1,2 \\
2003 Jun 06 & 2452796.80  &   +30.4      &  0.8-2.5   & UK-1,2 \\
\hline
\end{tabular} \\
\label{t:speclog:nir}
$^a$TNG = TNG + NICS, UK-1/2 = UKIRT + CGS4/UIST
\end{table}

\begin{figure*}[!ht]
 \centering
 \includegraphics*[width=17cm]{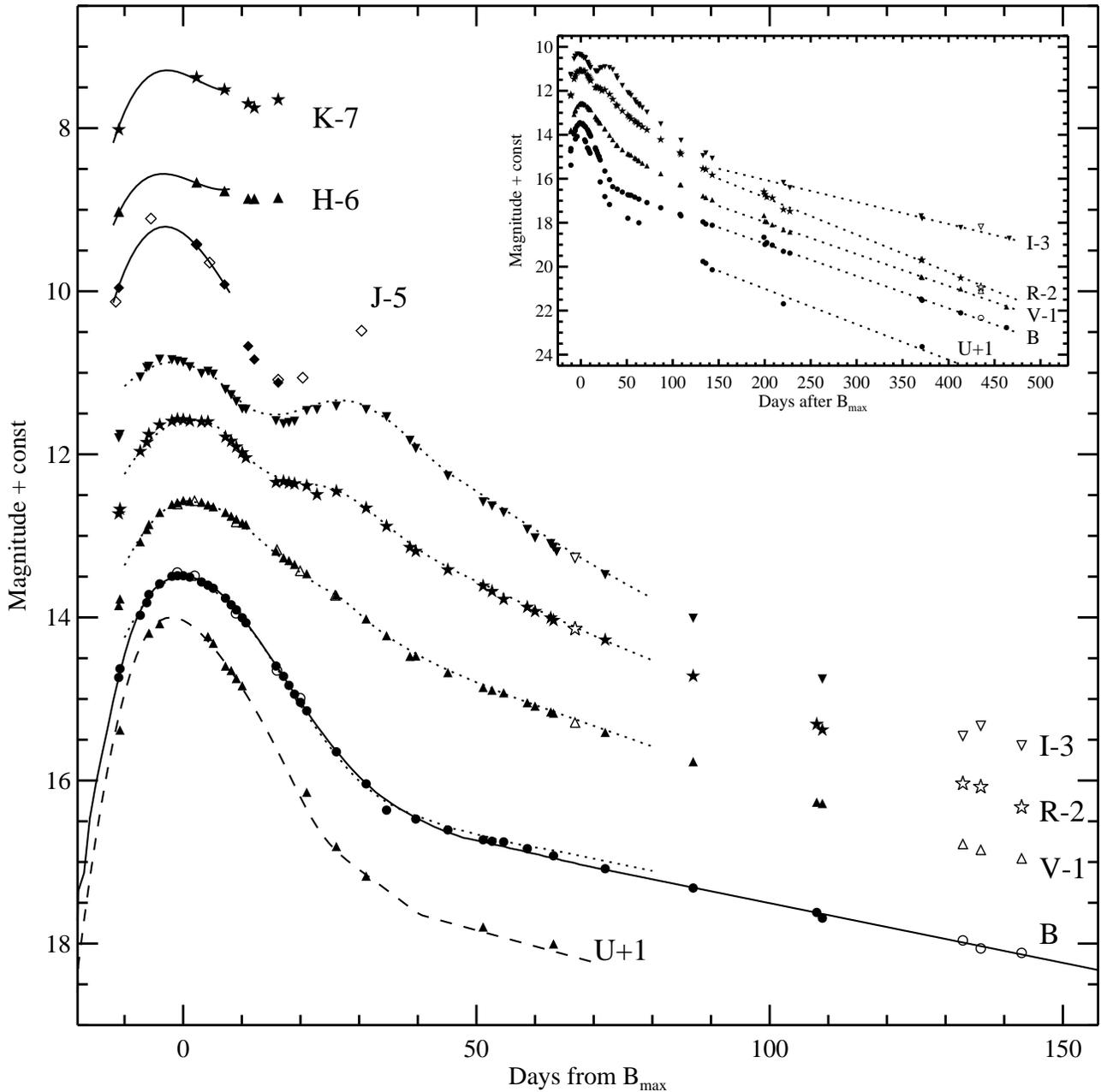}
 \caption{$UBVRIJHK$ light curves of \object{SN\,2003du}. The error bars are not plotted
  because they are typically smaller than the plot symbols.
  For the $BVRI$ bands the filled and the open symbols are the 
  S-corrected and non-S-corrected photometry, respectively. 
  The open symbols for $J$ band are synthetic photometry from the 
  combined optical-NIR spectra. Overplotted are the $B$-band template  of 
  \citet{nug2002},  the $JHK$ templates from \citet{kri_ir_temp}  (solid lines),
  as well as our $U$-band template derived from \citet{jha44} data (dashed line).
   The dotted lines are a light curve template with $\Delta m_{15}=1.02$ 
  calculated as described in \citet{prieto} using a program provided by the 
  authors.   {\bf Inset:} The full light curves. The late-time $HST$ data from 
  \citet{leo_du} are also
  shown with the open symbols. The linear fits to the late-time photometry are also shown.}
 \label{f:lc}
 \end{figure*}

The $UBVRIJHK$ light curves (LCs) of \object{SN\,2003du} are shown in
Fig.\,\ref{f:lc}.   The light curves
morphology resemble that of a normal SN Ia with a well-pronounced
secondary maximum in the $I$-band and a shoulder 
in the $R$-band. The $J$-band also shows a strong rise towards a secondary
maximum. Comparison with the photometry of \citet{leo_du}
and \citet{anu_du} reveals fairly good
consistency. However, systematic differences between the data sets do
exist  and our photometry is generally brighter. 
This is probably due to the differences in the comparison
star calibrations, as well as to the fact that our photometry was
S-corrected, unlike those of \citet{leo_du} and \citet{anu_du}.
 To estimate the 
differences we fitted a smoothing spline function 
to our data and computed the mean difference and its standard deviation from  
\citet{leo_du} and \citet{anu_du} photometry.
The difference slightly varies with the SN phase. Up to 30  days after maximum
light the mean differences and standard deviations in $BVRI$ are, respectively, 
$0.068\pm0.030$, $0.046\pm0.029$, $0.047\pm0.022$ and
$0.042\pm0.037$ mag with \citet{anu_du} and $0.026\pm0.025$, $0.026\pm0.015$,
$0.014\pm0.015$ and $0.071\pm0.030$ mag with \citet{leo_du}. The difference with
the  \citet{anu_du} $U$-band photometry is $0.007\pm0.085$ mag.

We fitted the $B$-band template of \citet{nug2002} to
the data to determine the $B$-band light curve parameters.  This
provided the time of $B$ maximum light
$t_{B_{\rm max}}$(JD)=2452766.38 (2003 May 6.88 UT), stretch factor
$s_B=0.988 \pm 0.003$ and peak magnitude $B_{\rm max}=13.49 \pm0.02$
mag. The peak $VRI$ magnitudes were estimated by fitting
low-order polynomials to the data around maximum, giving $V_{\rm
max}=13.57\pm0.02$, $R_{\rm max}=13.57\pm0.02$ and $I_{\rm
max}=13.83\pm0.02$ mag. The $U$-band maximum was estimated by fitting
our own template derived from the SNe published by \citet{jha44}: 
$U_{\rm max}=13.00\pm0.05$ mag.  The optical
photometric coverage around 15 days after $B_{\rm max}$ is rather
sparse. However, the $B$-band template matches the observed photometry
very well, thus we are able to use this to determine the decline rate
parameter.  We find $\Delta m_{15}=1.02\pm0.05$. $BVRI$ template light curves 
with $\Delta m_{15}=1.02$ were also generated using the data and the method
described by \citet{prieto}. These light curves are also shown in 
Fig.\,\ref{f:lc}, shifted to match \object{SN\,2003du} peak magnitudes. 
 The resemblance between \object{SN\,2003du} light curves and the templates is
excellent.

 \begin{figure*}[!t]
 \centering
 \includegraphics*[width=15cm]{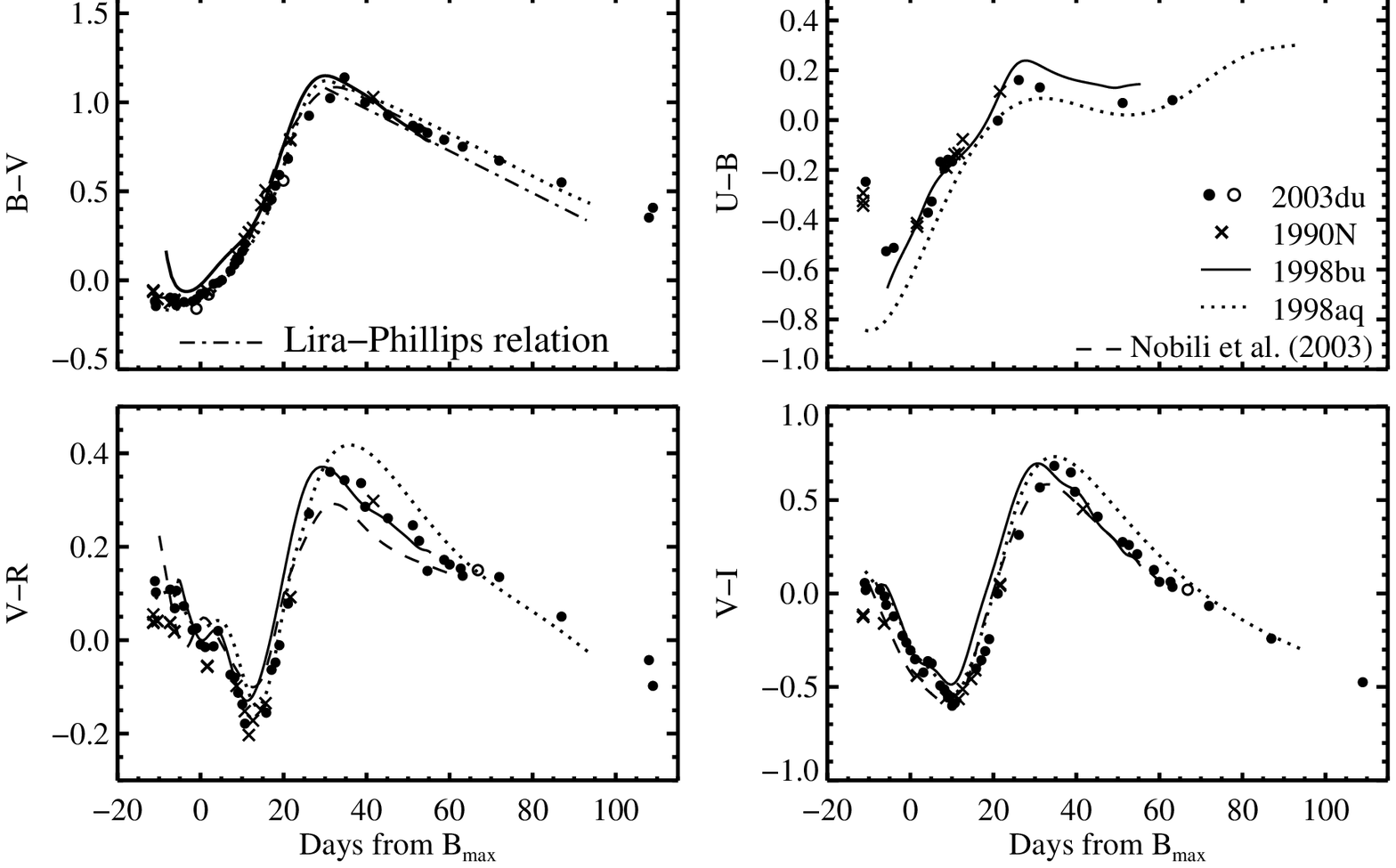}
 \caption{The evolution of the optical color indices of \object{SN 2003du}, compared
  with those of other well-observed SNe\,Ia. When necessary the colors were
   de-reddened with the appropriate $E(B-V)$.  The \citet{nob} $B-V$, $V-R$ and $V-I$ templates
  are also shown.}
  \label{f:ci}
 \end{figure*}

 The NIR templates
from \citet{kri_ir_temp} were fitted to the first
three $JHK$ photometric points (Fig.\,\ref{f:lc}) to estimate the peak magnitudes:
$J_{\rm max}=14.21$, $H_{\rm max}=14.56$ and $K_{\rm max}=14.29$ mag.  
The rms around the fits are fairly small 0.03, 0.02 and 0.04 mag, respectively,
 but 
because the LCs are undersampled
the uncertainties in the peak magnitudes should exceed these values.
To derive the templates, \citet{kri_ir_temp} 
fitted third-order polynomials to the photometry of a number of SNe. 
The rms around the fits are 0.062, 0.080 and 0.075 mag for $J$, $H$ 
and $K$, respectively. These numbers were added in quadrature to the rms 
around the fits to the \object{SN\,2003du} data to obtain the uncertainties of the 
$JHK$ peak magnitudes, 0.07, 0.08 and 0.09 mag, respectively.

The entire light curves are shown in the inset of Fig.\,\ref{f:lc}.
The late-time $HST$ data from \citet{leo_du} are also
shown (open symbols); these are consistent with our ground based photometry.
After $\sim+180$ days the magnitudes of \object{SN\,2003du} decline
linearly,  following the expected form of an exponential radioactive decay chain.
The decline rates in magnitudes per 100
days in $UBVRI$-bands (as determined by linear least-squares fitting) are
$1.62\,\pm0.12$, $1.47\,\pm0.02$, $1.46\,\pm0.02$, $1.70\,\pm0.06$ and
$1.00\,\pm0.03$, respectively. 
The decline rates in the
$B$- and $V$-bands are virtually the same.
The $I$-band decline on the other hand is much slower than in the other bands. 
 Many other normal SNe\,Ia \citep[e.g.,][]{lair} and the peculiar 
\object{SN\,2000cx} \citep{jesper} also show similar behavior.

\subsection{Reddening in the host galaxy}

Figure\,\ref{f:ci} shows that the time evolution of the color indices (CIs) of
\object{SN\,2003du} closely follows the reddening corrected CIs of
normal SNe such as \object{1990N}, \object{1998aq}, and \object{1998bu}, as well as 
the \citet{nob} templates. This implies that
\object{SN\,2003du} was probably not reddened within its host galaxy.
Nevertheless,  the reddening in the host galaxy was estimated with three
different methods. The CIs of \object{SN\,2003du} were first corrected for the small Milky Way 
reddening of $E(B-V)=0.01$ \citep{ebv} assuming $R_V=3.1$.

i) \citet{phil99} use the observed $B_{max}-V_{max}$
 and $V_{max}-I_{max}$ indices, and the evolution of $B-V$ between 30
 and 90 days after maximum to derive $E(B-V)$. The first two
 quantities are weak functions of $\Delta m_{15}$. The time evolution of
 $B-V$ (known as the Lira relation) seems to hold for 
 the majority of SNe\,Ia \citep{phil99,jha_mcls2002}.  
 Following \citet{phil99}, for \object{SN\,2003du} we obtain
 $E(B-V)_{max}=-0.01\,\pm0.04$, $E(V-I)_{max}=0.07\,\pm0.05$ and
 $E(B-V)_{tail}=0.05\,\pm0.07$. The errors indicate the intrinsic accuracy
 of the three methods as given in \citet{phil99},
 viz. 0.03, 0.04 and 0.05, added in quadrature to the uncertainties of
 the observed CIs. Note that the $B-V$ evolution of \object{SN\,2003du} has a
 different slope from that of the Lira relation, leading to rather a
 large scatter of $\sim0.07$ mag.  We averaged the above estimates of
 $E(B-V)_{max}$, 0.8$\times E(V-I)_{max}$\footnote{the factor 0.8 serves to
convert $E(V-I)$ to $E(B-V)$ assuming the standard Milky Way 
extinction law with $R_V=3.1$.} and $E(B-V)_{tail}$ weighted by their respective uncertainties 
to obtain the final reddening estimate: $E(B-V)=0.027\,\pm0.026$.

ii) \citet{cmag} introduced a novel method, CMAGIC, to
estimate the brightness and the reddening of SNe\,Ia. It is based on
the observation that between 5-10 to 30-35 days after maximum the $B$
magnitude is a linear function of $B-V$ with a fairly uniform
slope. Applying this method, we obtain $E(B-V)=0.00\,\pm0.05$. 

 \begin{figure}[!t]
 \centering
 \includegraphics*[width=8cm]{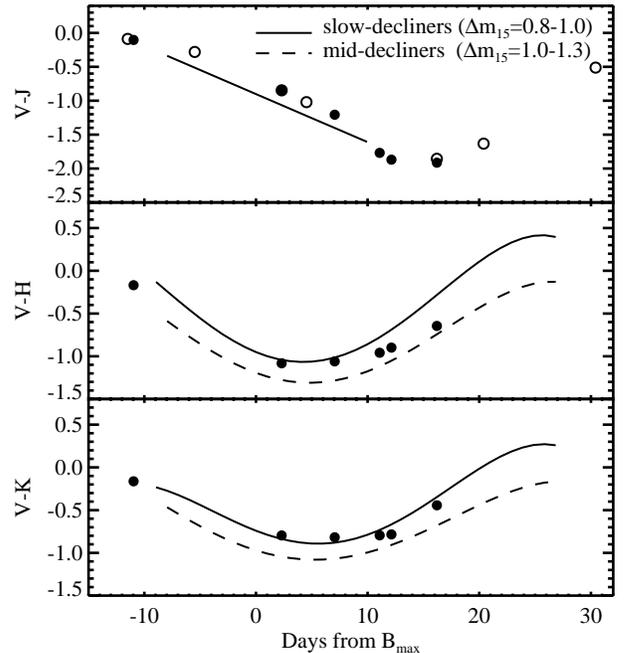}
 \caption{$V-(JHK)$ color indices of \object{SN\,2003du}. The
  unreddened loci for mid- and slow-declining SNe of
  \citet{kri_ir_temp} are overplotted. The open symbols are
  estimates based on synthetic photometry from the combined
  optical-NIR spectra.}
 \label{f:vnir}
 \end{figure}

iii) \citet{kri_00,kri_01,kri_ir_temp} have shown 
that the intrinsic $V-(JHK)$ CIs of SNe\,Ia are very uniform and 
can be used to estimate
the reddening of the host galaxy. Figure\,\ref{f:vnir} shows the
$V-(JHK)$ CIs of \object{SN\,2003du} overplotted with the unreddened
loci for mid- ($\Delta m_{15}=1.0-1.3$) and slow-declining SNe 
($\Delta m_{15}=0.8-1.0$) of \citet{kri_ir_temp}. Most of the $V-(HK)$ data of
\object{SN\,2003du} fall between the two loci.  This is consistent with the
fact that its $\Delta m_{15}=1.02$ lies between these two groups of
SNe\,Ia. Although the $V-J$ CI is slightly redder than the locus,
overall the $V-(JHK)$ CIs of \object{SN\,2003du} suggest little
reddening.

Combining the results of the three estimates we conclude that
\object{SN\,2003du} suffered negligible reddening within the
host galaxy. The main parameters of \object{SN\,2003du} that 
we derived from photometry are 
summarized in Table\,\ref{t:03dupars}.

\subsection{Spectroscopy}

 \begin{figure*}[!t]
 \centering
 \includegraphics*[width=14cm]{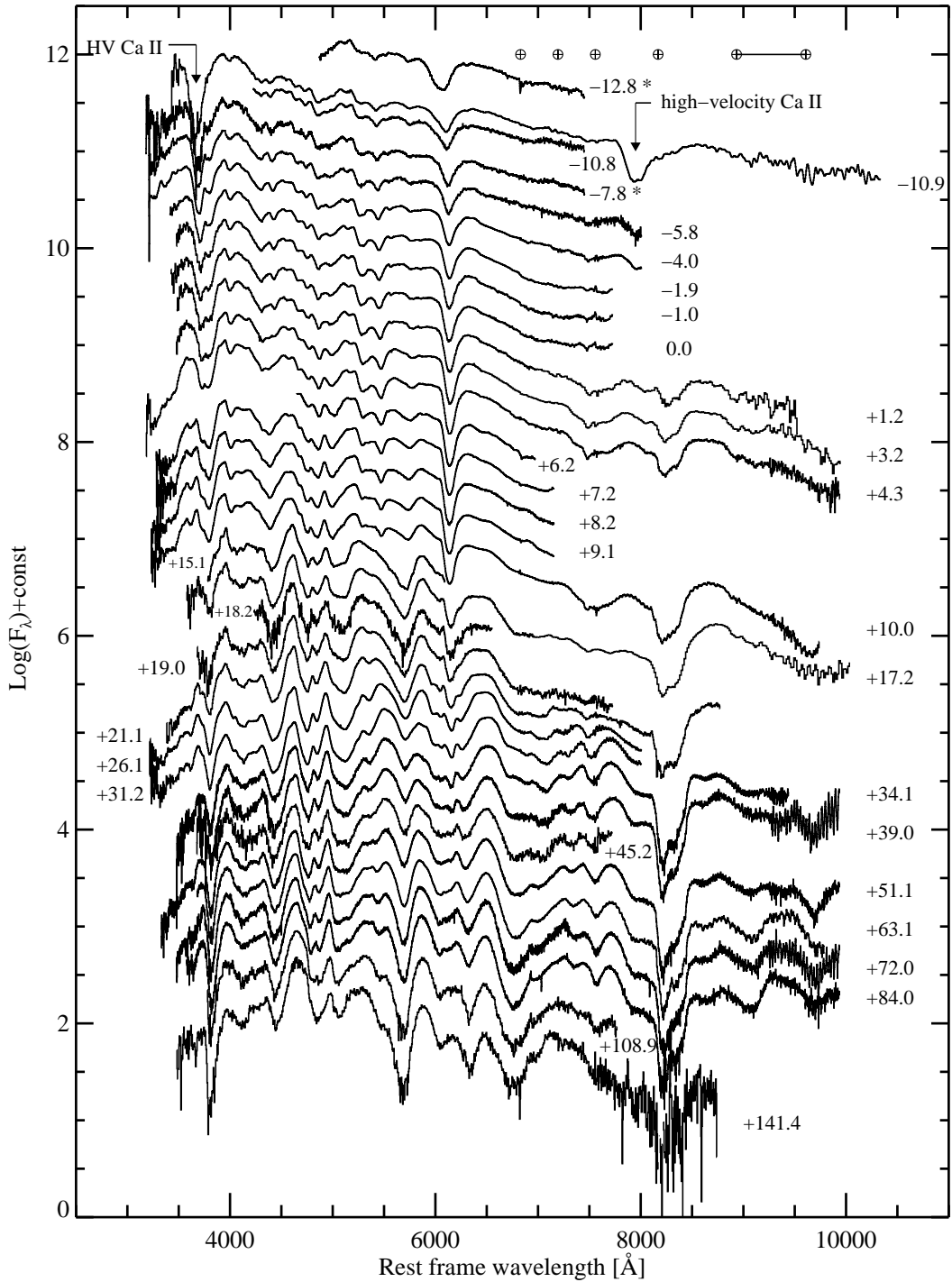}
 \caption{Evolution of the optical spectra of \object{SN 2003du}. The
 spectra marked with an asterisk were slightly smoothed (see text for
 details).  The noticeable telluric features are marked with Earth
 symbols; the connected symbols mark the region of strong telluric
 absorption.}
 \label{f:opt-sp}
 \end{figure*}

\begin{figure}[!t]
 \centering
 \includegraphics*[width=8cm]{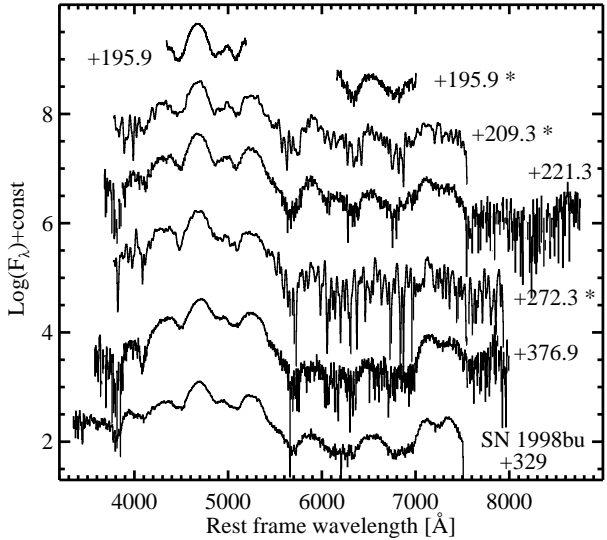}
 \caption{Nebular spectra of \object{SN 2003du}. The
 spectra marked with an asterisk were slightly smoothed. A
 nebular spectrum of \object{SN\,1998bu} is shown for comparison.}
 \label{f:opt-sp1}
 \end{figure}

Our collection of optical spectra of \object{SN\,2003du} is shown in
Figs.\,\ref{f:opt-sp} and \ref{f:opt-sp1}. The spectra marked with 
an asterisk have been smoothed 
using the {\it \'a trous} wavelet transform \citep{atrous}.
The optical spectral evolution of \object{SN\,2003du} is that
of a normal SN\,Ia. In the earliest spectrum at $-13$ days 
the \ion{Si}{ii}\,$\lambda$6355 line is strong and broad 
 ($\sim10000$\,km\,s$^{-1}$ full-width at half-depth), and the 
\ion{S}{ii}\,$\lambda$5454 and $\lambda$5640 lines are well developed.
 In the $-11$ day spectrum the \ion{Ca}{ii} H\&K and the IR triplet 
lines are also very strong.
In all the spectra until one week after maximum light,
\ion{Si}{ii}\,$\lambda$4129 and $\lambda$5972 lines are clearly visible.
\ion{Mg}{ii}\,$\lambda$4481, \ion{Si}{iii}\,$\lambda\lambda$4553,4568 and the blend of
\ion{Fe}{ii}, \ion{Si}{ii} and \ion{S}{ii} lines around
4500--5000\,\AA\ are also prominent. A few days after $B_{\rm max}$ 
the spectrum starts to be dominated by Fe group elements and gradually 
evolves into a nebular spectrum. 

 The ratio between the depth of the \ion{Si}{ii}\,$\lambda$5972 and $\lambda$6355 lines,  
$\mathcal{R}($\ion{Si}{ii}$)$ \citep{nugent}, at maximum is
$\mathcal{R}($\ion{Si}{ii}$)=0.22\,\pm0.02$, typical for
normal SN\,Ia. $\mathcal{R}($\ion{Si}{ii}$)$ does not change
significantly in the pre-maximum spectra, remaining at $\sim0.2$.

\begin{table}
\caption{Main photometric parameters of \object{SN\,2003du} from this work.}%
\begin{tabular}{@{}c|c@{}}
\hline
\hline
$t_{B_{\rm max}}$ [JD] &    $2452766.38\pm0.50$ \\
$t_{B_{\rm max}}$ (UT date) &  May 6.88, 2003  \\
 $B$-band stretch, $s_B$ & 0.988$\pm0.003$ \\
 $B$-band decline rate, $\Delta m_{15}$ & 1.02$\pm0.05$ \\
\hline
Peak magnitudes & $U=13.00\pm0.05$\hspace{0.2cm} $B=13.49\pm0.02$ \\
 &  $V=13.57\pm0.02$\hspace{0.2cm} $R=13.57\pm0.02$  \\
 &  $I=13.83\pm0.02$\hspace{0.2cm} $J=14.21\pm0.07$  \\
 & $H=14.56\pm0.08$\hspace{0.2cm} $K=14.29\pm0.09$ \\
\hline
Late-time decline &   $\gamma_U=1.62\pm0.12$\hspace{0.2cm}  $\gamma_B=1.47\pm0.02$ \\
rate $\gamma$  [mag/100 days] &  $\gamma_V=1.46\pm0.02$\hspace{0.2cm} $\gamma_R=1.70\pm0.06$  \\
& $\gamma_I=1.00\pm0.03$\hspace{0.2cm} $\gamma_{\rm Bol}=1.40\pm0.01$ \\
\hline
E$(B-V)_{\rm host}$  & 0.00\,$\pm0.05$ \\
\hline
\end{tabular}
\label{t:03dupars}
\end{table}

\begin{figure}[!t]
 \centering
 \includegraphics*[width=8.3cm]{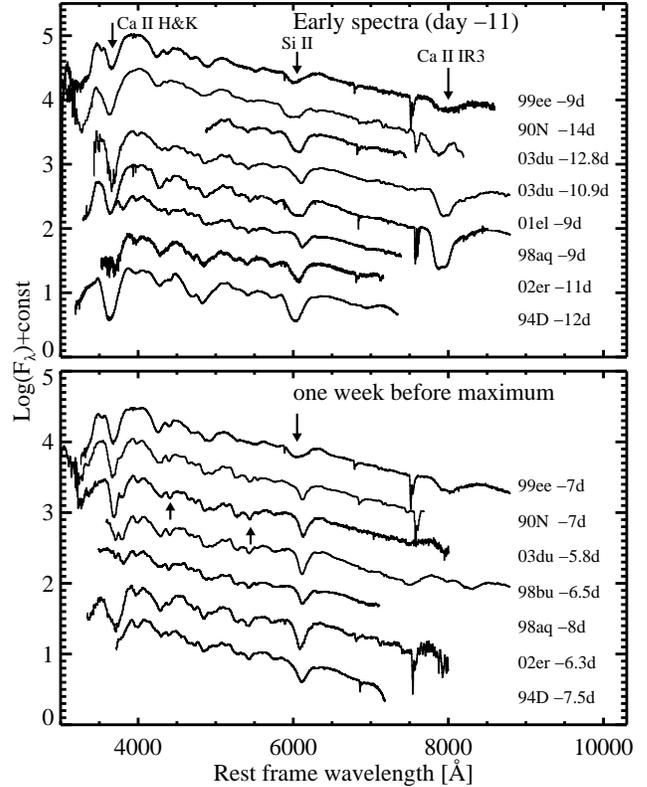}
 \caption{Comparison of optical spectra of normal SNe\,Ia at two
        pre-maximum epochs. 
        The arrows in the lower panel
        mark the lines whose velocities have been measured.}
 \label{f:sp-comp}
 \end{figure}

\begin{figure}[!h]
 \centering
 \includegraphics*[width=5cm]{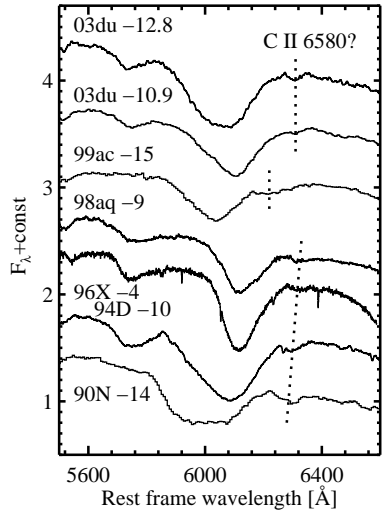}
 \caption{Early spectra of \object{SN\,2003du} and several other SNe\,Ia 
  zoomed at the \ion{Si}{ii}\,6355 line.  The dotted lines mark the weak 
  absorption features that may be due to \ion{C}{ii}\,$\lambda$6580.}
 \label{f:comp_c2}
 \end{figure}

In Fig.\,\ref{f:sp-comp} three of the pre-maximum spectra of \object{SN\,2003du} 
are compared with spectra of other normal SNe\,Ia observed at similar epochs and
appropriately de-reddened. For this and other comparison plots we
use published optical spectra of \object{SN\,1994D} 
\citep{nando94d,fil_araa,meikle96}, \object{SN\,1990N} \citep{bruno90n}, 
\object{SN\,1996X} \citep{salvo96x}, 
\object{SN\,1999ee}  \citep{mario99ee}, \object{SN\,1998aq}  \citep{branch_98aq}, 
\object{SN\,1998bu} \citep{jha98bu,her98bu}, 
\object{SN\,2002er} \citep{sp02er}, \object{SN\,2001el} \citep{wang01el,seppo_01el} and 
 \object{SN\,2005cg} \citep{05cg}.
The spectra at about 10 days before maximum
show significant differences. The spectra have not been taken at exactly the
same phase and the rapid spectral evolution at such early phases may partly be 
responsible for the differences. However, most of the differences 
are most likely intrinsic. It worths noting that the 
weak feature at $\sim$6300\,\AA\  that is visible in the two  earliest 
spectra of \object{SN\,2003du}    is present 
in other SNe\,Ia as well (Fig.\,\ref{f:comp_c2})
and has been attributed  to \ion{C}{ii}\,$\lambda$5860 
\citep{mazzali_90n,branch_98aq,gabri99aa,gabri99ac}.  
At one week before maximum the spectra 
are more similar to each other. 
It is interesting to note that at these
epochs the largest differences between the SNe are seen in the
strengths and profiles of the \ion{Si}{ii}\,$\lambda$6355, \ion{Ca}{ii} H\&K
and \ion{Ca}{ii}\,IR3 lines.
Starting from one week before maximum 
the spectra of most SNe\,Ia are very homogeneous.

\begin{figure}[!t]
 \includegraphics*[width=8.8cm]{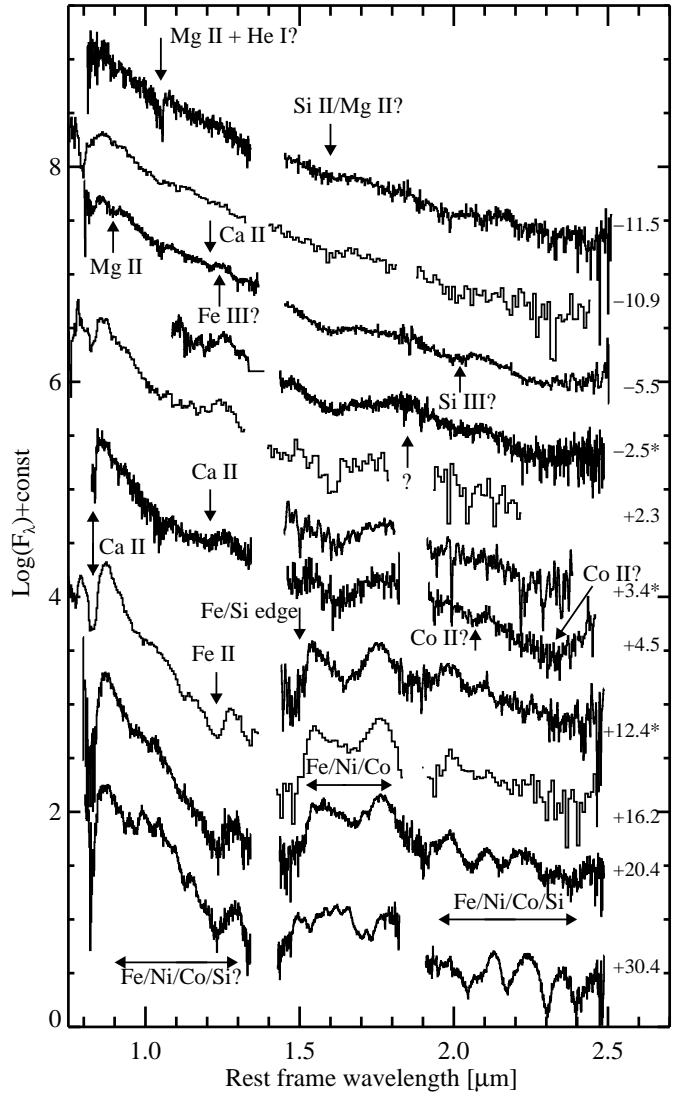}
 \caption{NIR spectral evolution of \object{SN 2003du}.
 The spectra marked with an asterisk have been smoothed 
 (only the $J$ band of the $-2.5$ days spectrum is smoothed).
}
 \label{f:ir-sp}
 \end{figure}

The NIR spectra of \object{SN\,2003du} are shown in
Fig.\,\ref{f:ir-sp}. 
 The earliest spectra at $-11.5$ and $-11$ days
are rather featureless with only hints at weak broad P-Cygni profiles.
The weak $\sim1.05$\,$\mu$m absorption could be due to \ion{Mg}{ii}\,$\lambda$10926 or
\ion{He}{i}\,$\lambda$10830 (or a combination of the two) 
\citep{meikle96,mazzali_he,branch_00cx,marion03}. 
The strength of this absorption in the earliest two spectra is quite different,
despite the fact that they have been taken only half a day apart. In the 
$-11.5$ days spectrum, however, the absorption is likely enhanced by a noise spike
due to the low instrument response at this wavelength.

In the day $-5.5$ spectrum an absorption due to \ion{Mg}{ii}\,$\lambda$9226 
\citep{marion03} is clearly seen. In the earlier 
IR spectra there are only hints of its
presence and it may be just detectable in the optical spectrum
at day $-11$. Our experiments with the SN spectral synthesis code SYNOW 
\citep[see for details, e.g.][]{branch_98aq} show however, that 
\ion{Si}{iii} and possibly \ion{Si}{ii} may contribute to the red part of this line.
No other features are detected in the 0.9-1.2\,$\mu$m 
spectral region. In particular, no \ion{C}{i} or \ion{O}{i}  lines are 
observed, in accordance with the findings of \citet{marion06}.
The absorption at  $\sim1.21$\,$\mu$m is due to \ion{Ca}{ii} 
according to \citet{wheeler98}, but the 
associated emission peak at $\sim1.24$\,$\mu$m was
attributed to \ion{Fe}{iii} by \citet{rudy} 
in \object{SN\,2000cx}.
The $1.6$\,$\mu$m absorption seen in the spectra until maximum light
is due to \ion{Si}{ii} with a
possible contribution from \ion{Mg}{ii}  \citep{wheeler98,marion03}. 
The broad features beyond $\sim1.7$\,$\mu$m lack clear identification. Possible contributors
are \ion{Si}{iii} at $\sim2$\,$\mu$m  \citep{wheeler98} and \ion{Co}{ii}
at $\sim2-2.05$\,$\mu$m and $\sim2.3$\,$\mu$m   \citep{marion03}.

By day $+12$, two
strong emission features at $\sim1.55$\,$\mu$m and $\sim1.75$\,$\mu$m
dominate the 1.4-1.8\,$\mu$m spectral region.
These two features  are
formed by blending of many \ion{Fe}{ii}, \ion{Co}{ii} and \ion{Ni}{ii}
emission lines \citep{wheeler98}.
Lines of \ion{Fe}{ii}, \ion{Co}{ii}, \ion{Ni}{ii} and \ion{Si}{ii}
dominate the spectral region beyond 2\,$\mu$m.
From day $+15$, a number of lines, with uncertain identifications 
also develop in the $J$ band.  One can also clearly see how a flux deficit at $\sim1.35$\,$\mu$m 
develops. This causes the very deep minimum observed in the $J$-band
light curves of most SNe\,Ia around 20 days after maximum.

\begin{figure}[!t]
 \includegraphics*[width=8.8cm]{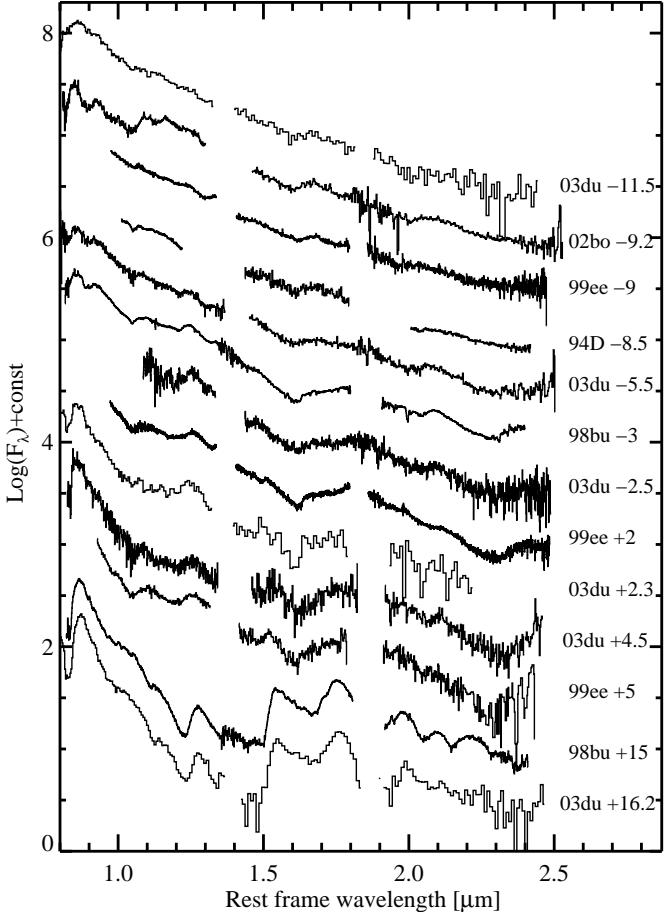}
 \caption{Comparison with NIR spectra of other normal SNe\,Ia.}
 \label{f:comp-ir}
 \end{figure}

Figure\,\ref{f:comp-ir} presents a comparison of several IR spectra of
\object{SN\,2003du} with those of other normal SNe: 
\object{SN\,1994D} \citep{meikle96},  
\object{SN\,1999ee} \citep{mario99ee},  \object{SN\,1998bu} \citep{jha98bu,her98bu,mario99ee}, 
and \object{SN\,2002bo} \citep{02bo}. Similarly to the
optical, the IR spectra of normal SNe taken at similar epochs are
very homogeneous,
 even the spectra taken 6--12 days
before maximum.  The only significant difference is in the $J$-band, 
where the \ion{Mg}{ii} lines of \object{SN\,2002bo} are
stronger compared to other SNe.

\subsection{Blueshifts of absorption-line minima}

We have measured the blueshifts of the absorption-line minima of 
\ion{Si}{ii}\,$\lambda$6355, \ion{S}{ii}\,$\lambda$5640 and
\ion{Si}{iii}\,$\lambda\lambda$4553,4568, which are thought to be relatively
un-blended \citep{branch_98aq}, by fitting a Gaussian to the line
absorption troughs. 
 In the rest of the paper we report the velocities that correspond 
to the measured blueshifts of the absorption-line minima (unless otherwise stated) 
and will refer to these as {\it velocities of the absorption lines}. 
By convention, these velocities are {\it negative} and we say that the velocity of a line 
{\it increases} from, e.g. $-20000$ to $-10000$\,km\,s$^{-1}$. The velocities inferred from 
an explosion model will be reported as {\it positive} numbers.

\begin{figure}[!t]
 \centering
 \includegraphics*[width=8.5cm]{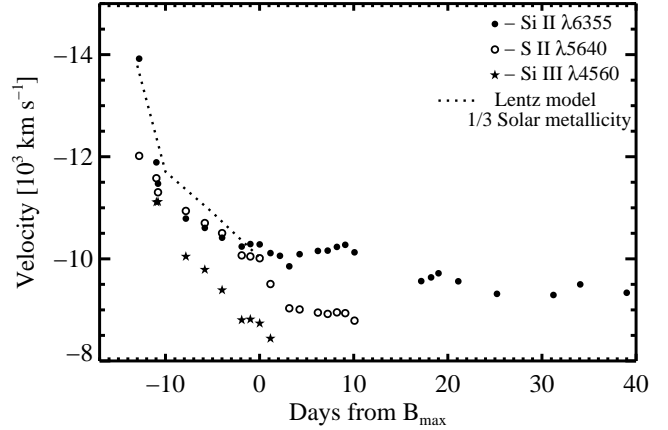}
 \caption{The evolution of the  velocity of the absorption lines
  of \object{SN\,2003du}.}
 \label{f:vel03du}
 \end{figure}

The velocities  of the \ion{Si}{ii}\,$\lambda$6355, \ion{S}{ii}\,$\lambda$5640 and
\ion{Si}{iii}\,$\lambda\lambda$4553,4568 lines are shown in Fig.\,\ref{f:vel03du} 
and it is evident  that the time evolution is very similar to that in 
other normal SNe\,Ia \citep[see, e.g.][]{ben_div}. 
The \ion{Si}{ii}\,$\lambda$6355 velocity initially  increases rapidly, but
7--5 days before maximum the increase rate
slows down and the velocity remains almost constant thereafter. 
The velocities of the \ion{S}{ii}\,$\lambda$5640 and \ion{Si}{iii}\,$\lambda\lambda$4553,4568  lines
 increase at nearly constant rate; however,
there is a hint that the \ion{S}{ii}\,$\lambda$5640 velocity remains
constant after maximum, similarly to \ion{Si}{ii}\,$\lambda$6355.
\citet{ben_div} measured a post-maximum velocity 
 increase rate of the \ion{Si}{ii}\,$\lambda$6355  line to be $\dot{v}= 31\pm5$ km\,s$^{-1}$d$^{-1}$ 
and classified \object{SN\,2003du} as a Low Velocity Gradient SN\,Ia, along with 
other normal and all overluminous SNe\,Ia.
It can be seen from Fig.\,1 in \citep{ben_div} that
before maximum the \ion{Si}{ii}\,$\lambda$6355  velocities of \object{SN\,2003du} are
systematically  higher by 500-2000\,km\,s$^{-1}$ compared to all other SNe. 

During the SN photospheric epochs the main source of continuum opacity 
at the optical wavelengths is electron scattering  and following 
\citet{jeff92} we adopt that the (continuum) photosphere is at electron scattering 
optical depth 2/3. However, the velocity gradient in the 
expanding SN ejecta causes many weak lines to overlap which gives rise to
strong  pseudo-continuum \citep[e.g.,][]{pau}, and the so-called expansion opacity 
\citep{karp,pe00} 
is an analytical description of this effect.
This expansion opacity may exceed electron scattering opacity 
by orders of magnitude.
The velocity of the pseudo-photosphere thus created 
is wavelength-dependent. 
Besides, strong absorption lines may form in a large 
volume above the  continuum photosphere. For these reasons, the
line  velocities
we measure most likely do not trace the 
velocity of the continuum photosphere and should be interpreted with caution.
\citet{lentz} have computed a grid of photospheric phase 
atmospheres of SNe\,Ia with different metallicities in the C+O layer and computed 
non-LTE synthetic spectra. 
It would be more reasonable for us to compare the  \ion{Si}{ii}\,$\lambda$6355 line velocities
 in \object{SN\,2003du}  with the measurements 
from the \citet{lentz} synthetic spectra.
 The time evolution is qualitatively similar and 
in Fig.\,\ref{f:vel03du} we also show 
the measurements for the 1/3 Solar metallicity models, 
which best follows the \object{SN\,2003du} \ion{Si}{ii}\,$\lambda$6355 line blueshift.

\citet{marion03} showed that the velocities of lines
in NIR spectra could be used to constrain the location of the transition region
between the layers of explosive carbon and oxygen burning, and
incomplete to complete silicon burning, and hence place constraints on
the explosion models.
We measured the  velocities of the blue edges of the absorptions at
$\sim0.9$\,$\mu$m and 1.05\,$\mu$m in our optical and IR spectra between $-11.5$
and $+4.5$ days. Both lines show constant velocities of
 $\sim-11000$\,km\,s$^{-1}$ and $\sim-13000$\,km\,s$^{-1}$, respectively, assuming that the
lines are formed by \ion{Mg}{ii}\,$\lambda$9226 and \ion{Mg}{ii}\,$\lambda$10926. The
constant velocity 
indicates that the  continuum photosphere is well beneath the
Mg-rich layers  \citep{meikle96}. The velocity of the 
sharp edge at $\sim1.55$\,$\mu$m in
the spectra between +10 and +20 days can be used to estimate the
transition between the layers of incomplete and complete silicon
burning.  We measure velocities $\leq-9800$\,km\,s$^{-1}$ which is similar to the 
results of \citet{marion03} and is also broadly consistent with 
their reference explosion model in which Si is completely consumed below 
$\sim8500$\,km\,s$^{-1}$. This ties in 
with the measurements of the  \ion{Si}{ii}\,$\lambda$6355 line velocity, which is always
 $\leq-9300$\,km\,s$^{-1}$.

\section{Discussion}

\subsection{The distance to \object{SN\,2003du}}

We have shown that \object{SN\,2003du} was a spectroscopically and
photometrically normal SN\,Ia, and furthermore that it was not reddened within its
host galaxy.  The distance to \object{UGC\,9391} has not been measured using direct techniques, and 
the only available information is from its recession velocity. 
The observed velocity is  1914\,km\,s$^{-1}$, which after correcting for 
the Local Group in-fall onto Virgo becomes  2195\,km\,s$^{-1}$ (from the LEDA database) or 
a distance modulus of $\mu=32.42$ mag on the scale of $H_0=72$\,km\,s$^{-1}$,Mpc$^{-1}$.

Recently, \citet{riess_98aq} calibrated the luminosities of
\object{SN\,1998aq} and \object{SN\,1994ae} by observing Cepheids in
their host galaxies with the {\it Hubble Space Telescope}. 
Including two other SNe\,Ia with Cepheid calibrated distances, 
they estimated the
absolute magnitude of a typical SN\,Ia to be $M_V=-19.20$\footnote{\citet{riess_98aq}
estimated $H_0=73$\,km\,s$^{-1}$,Mpc$^{-1}$ and $M_V=-19.17$, and 
we converted their $M_V$ to the scale of $H_0=72$\,km\,s$^{-1}$,Mpc$^{-1}$} 
$\pm0.10$(statistical)\,$\pm0.115$(systematic) mag. 
\citet{meikle00} and \citet{kri_04a,kri_04}
presented evidence that SNe\,Ia are  standard candles in the NIR
and that no correction for the light curve shape is needed for SNe
with $\Delta m_{15}<1.7$ mag. \citet{kri_04}
derived the following absolute peak $JHK$ magnitudes for $H_0=72$\,km\,s$^{-1}$,Mpc$^{-1}$:
$-18.61$,$-18.28$ and $-18.44$ mag all with statistical uncertainty of
$\sim0.03$ mag. The systematic uncertainty of $M_V$ is mostly due to the 0.1 mag
uncertainty in the distance to the Large Magellanic Cloud (LMC) and hence
it also affects the NIR absolute magnitudes and the distance modulus 
derived from the host galaxy recession velocity (through $H_0$).
The light curve decline rate parameter $\Delta m_{15}=1.02\pm0.05$ 
and the normal spectral evolution suggest that 
\object{SN\,2003du} is very  similar to normal SNe\,Ia. If one assumes that 
\object{SN\,2003du}  had the above-mentioned absolute $VJHK$ magnitudes,  
a distance modulus of $\mu=32.79\,\pm0.04$ (or a radial velocity of
$\sim2600$\,km\,s$^{-1}$ with $H_0=72$\,km\,s$^{-1}$,Mpc$^{-1}$) is obtained
\footnote{The absolute $JHK$ magnitudes of \citet{meikle00}
are by 0.4 mag brighter than those of 
\citet{kri_04}.  The distance moduli derived with the values from
the latter paper are consistent with the estimates of absolute $V$ 
magnitude from \citet{riess_98aq}; we therefore 
 adopt the \citet{kri_04} values.}.
This estimate is the  average of the four individual estimates
 weighted by their statistical uncertainties, i.e. the errors of \object{SN\,2003du} peak $VJHK$ magnitudes
added in quadrature to the statistical uncertainties of the absolute magnitudes. 

The difference between the two distance moduli is 0.37
mag  (it will further increase
if the \citealt{meikle00} absolute NIR magnitudes are used) and indicates that 
\object{SN\,2003du} was fainter than the average of SNe with $\Delta m_{15}=1.02$. 
The $1\sigma$ dispersion of SNe\,Ia absolute magnitudes in both, optical and IR, 
is $\sim0.15$ mag \citep[e.g., ][]{phil99,kri_04}. The uniformity and the small dispersion 
of the $V-[JHK]$ colors of SNe\,Ia \citep{kri_ir_temp} 
indicates that the intrinsic scatter in the $VJHK$ bands is {\it correlated}, and so cannot 
be reduced by averaging observations in different bands.
 Therefore, the distance modulus we estimate, $\mu=32.79\,\pm0.04$ mag, has an additional 
$\sim0.15$ mag uncertainty from the intrinsic dispersion of SNe\,Ia luminosity.  
 The fact that \object{SN\,2003du} is 0.37 mag fainter than expected for SNe with 
$\Delta m_{15}\sim1.02$ may thus be due to the intrinsic scatter ($2.5\sigma$ from of the mean).
It is also possible that \object{UGC\,9391} may not be in the undisturbed 
Hubble flow: if it has $v_r=2600$\,km\,s$^{-1}$ and a 
peculiar velocity component of $\sim400$\,km\,s$^{-1}$ toward the Earth, it may  seem
closer than it really is.  \object{UGC\,9391} is nearly face-on 
and the contribution of the galaxy rotation should be small.

\subsection{The bolometric light curve}

In order to compute the $uvoir$ "bolometric" light curve 
 (i.e. the flux within the 0.2-2.4$\mu$m interval) of
\object{SN\,2003du} 
we proceeded as follows.  First, our $U$-band
template LC was fitted to the $U$ photometry in order to estimate the
$U$ magnitudes when only $BVRI$ were available. The magnitudes were
corrected for the small Galactic reddening and transformed to flux
densities using the absolute calibration of the $UBVRI$ system by
\citet{bes_cal}. A cubic spline was fitted through the
data points and the resulting fit was integrated numerically over the
interval 3500-9000\,\AA.

Most of the early-time SN\,Ia luminosity is emitted at optical wavelengths, 
however, a non-negligible correction 
for the flux emitted outside the optical wavelengths is also needed 
\citep[see, e.g.][]{sun96}. The flux emitted beyond 
9000\,\AA\ was estimated by integrating the combined optical-NIR
spectra of \object{SN\,2003du}.  The filled circles in
Fig.\,\ref{f:bol}a show the time evolution of the ratio of the flux emitted
in the 9000-24000\,\AA\ range to that emitted within 3500-9000\,\AA.
\citet{sun96} finds that at $+80$ days
less than 10\% of the flux is emitted in the IR. We estimate from 
the photometry of \object{SN\,2001el} \citep{kri_01el}
that the contribution of the IR flux
is $\sim25$\% and $\sim15$\% at $+28$ and $+64$ days,
respectively. This is consistent with our estimates for
\object{SN\,2003du} and the findings of \citet{sun96}, and indicates 
that the contribution of the IR flux decreases roughly linearly between
days $+30$ and $+80$.

 As there are no UV spectra of \object{SN\,2003du} observed, we used UV spectra
of other SNe Ia to estimate the contribution of the UV flux.  These
comprised combined de-reddened UV-optical spectra of
\object{SN\,1990N} at $-14$ and $-7$ days \citep{bruno90n}, \object{SN\,1989B} at $-5$ 
(\citealt{wells} and UV spectra from the $IUE$ archive),
\object{SN\,1981B} \citep{branch_81b},
\object{SN\,1992A} at $+5$, $+9$ and $+17$ \citep{kir92a}, and \object{SN\,2001el} between $+30$ and $+66$
(from $HST$ archive).  For spectra that did not cover the full 
2000--9000\,\AA\ range we extrapolated to
9000\,\AA\ using spectra of \object{SN\,2003du}. The spectra of \object{SN\,2001el} were
linearly extrapolated from $\sim2900$\,\AA\ down to 2000\,\AA\ assuming that 
the flux approached zero at 1000\,\AA. 
In Fig.\,\ref{f:bol}a we show the ratios of the fluxes
in the 2000--3500\,\AA\ range to those in the 3500--9000\,\AA\ range
(open symbols). 

The total contribution of the UV and IR fluxes is
plotted as a dashed line in Fig.\,\ref{f:bol}a, and one can see the
particularly large corrections needed   before the $B$-band maximum and 
around the secondary $I$-band maximum. 
Beyond +80 days we assumed a constant IR contribution of 10\% and
that the UV contribution decreases linearly from 5\% at +80 days to zero at +500 days.
This correction was applied to the optical fluxes to
derive the $uvoir$ fluxes of \object{SN\,2003du}. These were 
converted to luminosity assuming a distance modulus $\mu=32.79$ mag.
The $uvoir$ "bolometric" light curve of \object{SN\,2003du} is shown in
Fig.\,\ref{f:bol}b. For comparison, we also show the bolometric light curve of \object{SN\,2005cf} 
\citep{pas05cf}, which is very similar to that of \object{SN\,2003du}.
The maximum $uvoir$ "bolometric" luminosity of \object{SN\,2003du} is
$1.35(\pm0.20)\times10^{43}$\,erg\,s$^{-1}$ at $\sim2$ days 
before the $B$-band maximum.  Using Arnett's Rule as formulated by 
\citet[][their Eq.\,7]{stritz_arnet}) we estimate
the amount of \element[][56]{Ni} synthesized during the explosion,
$M_{^{56}{\rm Ni}}=0.68\,\pm0.14\,M_{\sun}$.  The error is a simple propagation 
of the uncertainty of the bolometric peak luminosity and the relation 
of \citet{stritz_arnet}. 
However, \citet{kh93} have shown that the simplifying 
assumptions made in the derivation of Arnett's rule may lead to errors as large as 50\%.
Combined with the uncertainty of the distance to \object{SN\,2003du}, 
clearly this estimation of the \element[][56]{Ni} mass is subject to large 
systematic uncertainty. Note, however, that \citet{stritz_prog} have analyzed 
a nebular spectrum and the optical
photometry of \object{SN\,2003du}, and derived 
$M_{^{56}{\rm Ni}}\simeq0.6\,M_{\sun}$, which is  in good agreement with our estimate. 
If one accepts a distance modulus of $\mu=32.42$ mag ($\sim30.4$ Mpc), then 
the estimated peak luminosity and $M_{^{56}{\rm Ni}}$ should be reduced by $\sim30$\%.

\subsection{Bolometric light curve modeling}

To further estimate the amount of \element[][56]{Ni} synthesized 
we modeled the bolometric light curve of \object{SN\,2003du} for 
both distance moduli $\mu=32.42$ and $\mu=32.79$ mag.
We used the Monte Carlo light curve code  described by 
\citet{capp97} and \citet{maz01}. Starting from an explosion model and
a given \element[][56]{Ni} content the code computes the transport and deposition 
of the $\gamma$-rays and the positrons generated by the decay chain
\element[][56]{Ni}$\rightarrow$\element[][56]{Co}$\rightarrow$\element[][56]{Fe} in a grey 
atmosphere. The optical photons that are generated by the thermalization of the 
energy carried by the $\gamma$-rays and the positrons are then followed as they
propagate through the SN ejecta. The optical opacity encountered by these
photons is again assumed to be grey and to depend primarily 
 on the relative abundance of iron-group elements. The opacity also decreases 
with time as $(t_{\rm d}/17)^{-3/2}$, $t_{\rm d}$ being the time since the
explosion in days, to mimic the effect of the decreasing temperature. For more details on
the adopted parametrization of the opacity see, e.g. \citet{maz01}.
This simple approximation works well 
\citep[e.g.][]{maz01} but an alternative view that the opacity 
depends primarily on temperature has been suggested \citep{kas_woo}. 
 In \citet{maz00_mc_comp} the  Monte Carlo code was compared 
with the results from the  radiation hydrodynamics code of \citet{hydro}, 
finding very good agreement.

We followed the approach of \citet{maz_pod}, who
assumed that stable Fe-group isotopes (e.g. \element[][54]{Fe}, \element[][58]{Ni})
 may be present not only in the innermost part of the ejecta 
($\leq0.2\,M_{\odot}$), but also in the 
\element[][56]{Ni} zone between $\sim0.2\,M_{\odot}$ and $\sim0.8\,M_{\odot}$.
\citet{maz_pod} suggested that the scatter of SNe\,Ia 
luminosity at a given $\Delta m_{15}$ may be reproduced by changing the ratio 
of the amount of radioactive \element[][56]{Ni} and the stable isotopes
in the \element[][56]{Ni} zone, while keeping the total mass of the Fe-group elements constant. 
This ratio may be sensitive, for example, to the metallicity of the progenitor white dwarf
\citep{timmes}. The SN\,Ia light curve width is mainly 
 determined by the opacity of the ejecta, which in turn is mostly determined by the total amount 
(stable and radioactive) of Fe-group elements synthesized, provided the temperature is above
$\sim10^4$\,K \citep[e.g.][]{kh93}. The peak luminosity on the
other hand is determined by the amount of \element[][56]{Ni}. Therefore, if the fraction
of stable Fe-group isotopes is varied within reasonable limits ($\sim 20$\%)
the temperature may not be affected significantly, and thus the opacity may be
effectively unchanged. This would lead to light curves with the same width, but different
luminosities.

 \begin{figure}[!t]
 \includegraphics*[width=8.8cm]{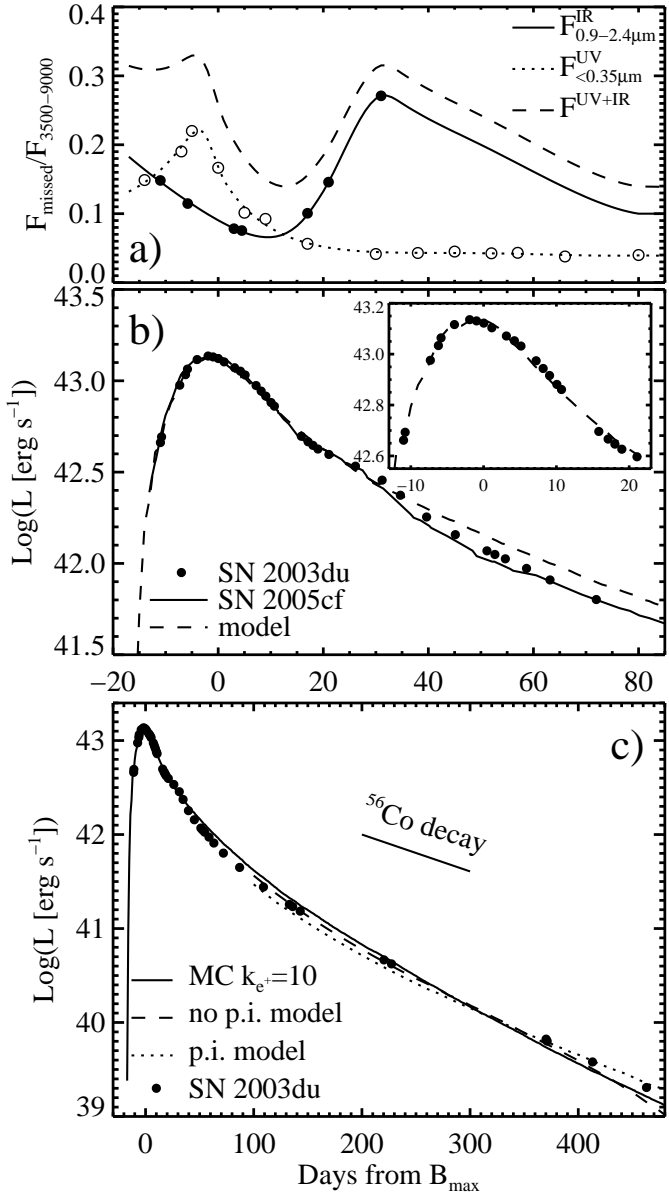}
 \caption{ {\bf a:} The ratio between the UV and IR fluxes to the flux within
 3500-9000\,\AA;  {\bf b:} The {\it uvoir} bolometric light curves of \object{SN 2003du}, 
 \object{SN 2006cf} and best model. The inset shows an expansion of the 
  \object{SN 2003du} light curve and model around maximum; {\bf c:} 
 The entire {\it uvoir} bolometric light curve with the models overplotted.}
 \label{f:bol}
 \end{figure}

As shown in Fig.\,\ref{f:bol}b, the $uvoir$ "bolometric" light curve of \object{SN\,2003du}
is remarkably similar to that of \object{SN\,2005cf} \citep{pas05cf} 
if $\mu=32.79$ mag is adopted. Therefore, a model similar to that adopted for 
\object{SN\,2005cf} can be used also to reproduce the light curve of \object{SN\,2003du}. 
In this case the best fit, shown in Fig.\,\ref{f:bol}b, is obtained for a model with 
0.69$M_{\odot}$ of \element[][56]{Ni}  and 0.42$M_{\odot}$ of stable 
Fe-group isotopes using the W7 explosion model \citep{no84} 
as an input. This estimate of the amount of \element[][56]{Ni} is in 
excellent agreement with the estimate derived above using Arnett's rule.
However, mixing out of a sufficient amount of \element[][56]{Ni} is necessary to 
reproduce the early rise of the light curve. This is a feature that is not 
present in one-dimensional explosion models, but is often inferred from SN data. 
 For example, for \object{SN\,2002bo}, using the abundance distribution and the amount of 
\element[][56]{Ni} mixed out as derived from an abundance tomography experiment 
\citep{ab_tom} gave a much better reproduction of the bolometric 
light curve. What is interpreted as mixing in one-dimensional models may be 
related to the presence of high velocity features \citep{mazzali99ee}, 
which affect the early spectra of \object{SN\,2003du} 
quite heavily.

If the true distance  modulus were $\mu=32.42$, the light curve could only be reproduced 
if the total mass  of iron group elements was the same as above (i.e. 1.11$M_{\odot}$) 
but the \element[][56]{Ni} content was $\sim0.45\,M_{\odot}$.
While this may still be a possibility, with such a low \element[][56]{Ni} mass 
(less than half of the total Fe-group content) it can be expected that the
heating by radioactive decay is not sufficient to keep the gas at a
sufficiently high temperature ($\sim 10^4$K) that the opacity is unchanged. At
lower temperatures, the opacity rapidly drops \citep{kh93},
and thus the light curve would not be as broad as observed. We therefore
suggest that a reasonable range of distances for \object{SN\,2003du} is between 
$\mu=32.7$ and 33.0 mag, implying a \element[][56]{Ni} mass between 0.6 and 
0.8$M_{\odot}$ for a total Fe-group elements mass of $\sim1.1M_{\odot}$.

 Roughly 200 days after maximum SN\,Ia ejecta become
transparent to the $\gamma$-rays and the main source of energy  is
the positrons produced by the decay of \element[][56]{Co}. 
If the positrons are fully trapped and deposit all their 
kinetic energy, the {\it true} bolometric LC should have a decline rate 
of $\sim1$ mag per 100 days. Larger decline rates are 
typically found in SNe\,Ia, and assuming that the optical flux follows the 
{\it true} bolometric flux, this is usually interpreted as evidence
for positron escape \citep[see, e.g., ][]{col80,capp97,rl98,milne99}. The $uvoir$ 
"bolometric" luminosity decline rate of \object{SN\,2003du} after 
200 days is 1.4 mag per 100 days. However, 
late-time NIR observations of few SNe\,Ia have recently been published
(\object{SN\,1998bu} -- \citealt{spy_98bu};
 \object{SN\,2000cx} -- \citealt{jesper}; 
 \object{SN\,2004S} -- \citealt{kri_04s}) and 
indicate that after 300--350 days the NIR luminosity does not decline but stays nearly 
constant.  The contribution of the NIR flux therefore increases with time and 
if accounted for may  lead to decline rates lower than the observed ones and
closer to the full positron trapping value.  \citet{03du_late_ir}
obtained late-time NIR spectra (1.1-1.8\,$\mu$m) and $H$-band photometry of \object{SN\,2003du}.
At +330 days \object{SN\,2003du} had an $H$ magnitude 
of 20.12\,$\pm0.17$ (Motohara et al. private communication) and we  calculate
the integrated flux across the $H$-band  to be
$\sim3$\% of the optical flux at that epoch.
The late-time NIR spectra of \object{SN\,2003du} indicate that the 
 integrated $J$ and $H$ band fluxes are nearly equal, implying that the 
contribution of the NIR flux is at least 6\%. If we adopt a 10\% NIR contribution at 
+330 days and assume that the total NIR {\it flux} did not change afterwards,
we obtain a decline rate of 1.2 mag per 100 days, which is still larger than the
full positron trapping value.

In Fig.\,\ref{f:bol}c 
we compare the $uvoir$  "bolometric" LC of \object{SN\,2003du}
with the two models presented by \citet{jesper}. 
The models are in the form of broadband 
$U$-to-$H$ magnitudes. For a consistent  comparison with 
\object{SN\,2003du} we  used only the $UBVRI$  model
fluxes to compute the model $uvoir$ LC in exactly the same way as for \object{SN\,2003du}. 
The models are generic, and have not been tuned 
to any particular SN. They have been computed with 0.6$M_\odot$ \element[][56]{Ni} 
and assume full positron trapping, and differ only in the treatment 
of the photoionization representing two extreme cases that the UV photons 
either escape or are fully redistributed to lower energies
(for more details see \citealt{jesper} and  references therein).
For a comparison with \object{SN\,2003du} the models were only 
re-scaled to a distance modulus $\mu=32.79$ mag, and yet they fit the absolute flux 
level of the LC of \object{SN\,2003du} quite well.  
It is evident from  Fig.\,\ref{f:bol}c that
a model  with an intermediate treatment of the photoionization could reproduce the 
\object{SN\,2003du} light curve.
Figure\,\ref{f:bol}c also shows a model computed with the Monte Carlo code 
using the best parameters we estimated above. Only the opacity to positrons 
$k_{\beta^+}$ was adjusted to fit the late-time decline rate 
\citep{capp97}.  The best value is 
$k_{\beta^+}=10$\,cm$^{-2}$\,g$^{-1}$, which is well within
the range of values found by \citet{capp97}.

Both late-time LC models we discussed are based on the 1D W7 explosion model
and do not  include  a contribution from magnetic fields. However, detailed calculations 
\citep{rl98,milne99}
show that the positron deposition rate is quite sensitive to the 
magnetic field configuration in the ejecta and the actual explosion model.
Clearly, to fully exploit the information in the bolometric LC a 
more detailed study is needed, but this is beyond the scope of this paper.

\subsection{Evolution of \ion{Si}{ii}\,$\lambda$6355, \ion{Ca}{ii} H\&K and IR triplet}

 \begin{figure}[!t]
 \includegraphics*[width=8.8cm]{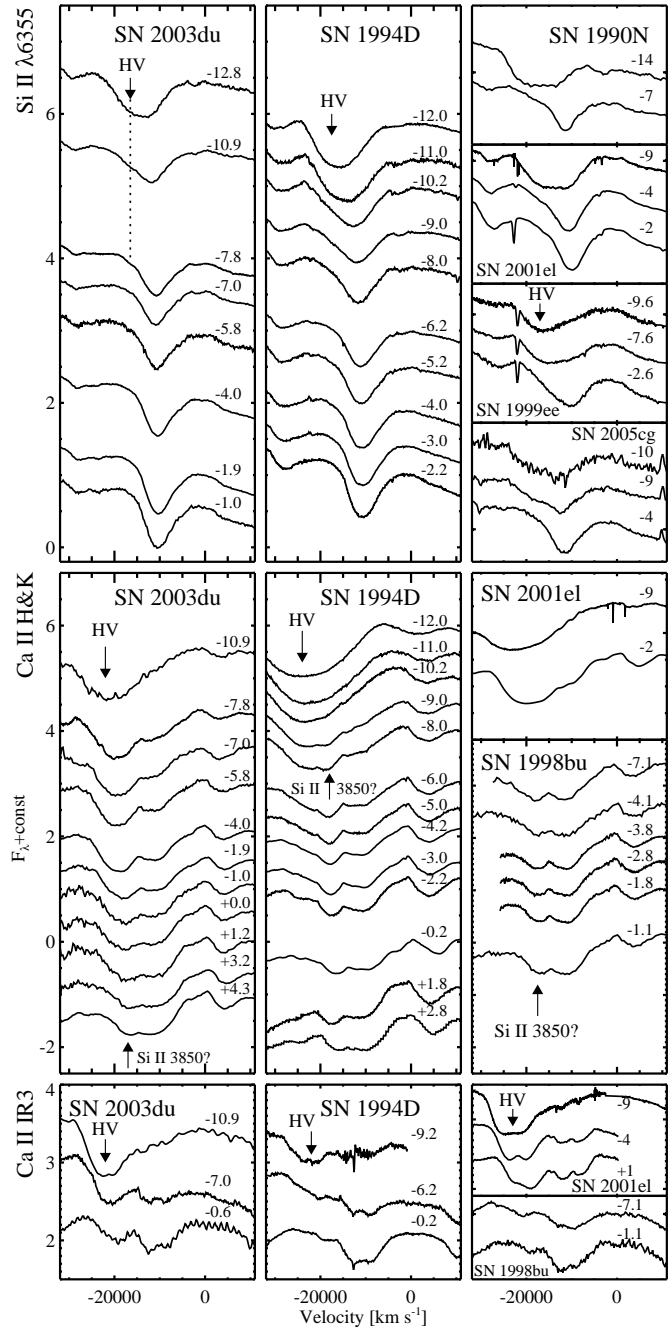}
 \caption{Comparison of the evolution of \ion{Si}{ii}\,$\lambda$6355,  
  \ion{Ca}{ii} H\&K and \ion{Ca}{ii} IR3  lines
 in \object{SN\,2003du} with those of other normal SNe\,Ia.}
 \label{f:evo}
 \end{figure}

Figure\,\ref{f:evo} shows the pre-maximum evolution of the 
absorption lines \ion{Si}{ii}\,$\lambda$6355, \ion{Ca}{ii} H\&K and \ion{Ca}{ii}~IR3 
in \object{SN\,2003du}  (here we also use a few spectra of \object{SN\,2003du} from 
 \citealt{ger_du} and \citealt{anu_du}) and other SNe\,Ia.
In the $-13$ days spectrum of  \object{SN\,2003du} the \ion{Si}{ii}\,$\lambda$6355 line 
is broad and rather symmetric. In the $-11$ day spectrum the line is
asymmetric and narrower, but around a week before maximum 
becomes symmetric again and the profile does not change much until maximum. 
The line evolution in \object{SN\,1994D} is very similar, but is delayed 
with respect to \object{SN\,2003du}: the $-13$ and $-11$ day spectra of 
\object{SN\,2003du} are most similar to those of \object{SN\,1994D} at 
$-11$ and $-9$ days.
Similar evolution is also observed in  \object{SN\,2001el}, \object{SN\,1990N},
 \object{SN\,1999ee} and  \object{SN\,2005cg}, but the pre-maximum coverage of
these SNe is rather sparse. Nevertheless, this profile 
evolution may be explained if the \ion{Si}{ii}\,$\lambda$6355 line
is a blend of two components. At $10-14$ days before maximum,
the strength of the two components should be nearly equal. The blue
component then decreases very rapidly, disappearing by $\sim7-5$ days
before maximum, while the red component increases in strength.
In \object{SN\,2003du}, the blue component was last seen in the 
$-7.8$ day spectrum as a weak feature on the blue wing of the line, and
in  \object{SN\,2001el} it may be still present in the $-2$ day spectrum.
The peculiar flat-bottom line shape in the early spectra of 
\object{SN\,2001el} and \object{SN\,1990N} is thus due to the blue component 
extending over a larger velocity interval compared to other SNe.
The $-9$ day spectrum of \object{SN\,1999ee} on the other hand, has a
stronger blue component such that the line is asymmetric with an extended {\it red} wing.
Note that \citet{mazzali_90n}  and \citet{mazzali99ee} find that 
a two-component model is needed to explain 
the peculiar \ion{Si}{ii}\,$\lambda$6355 line shape in \object{SN\,1990N} and \object{SN\,1999ee},
the high-velocity (HV) component being carbon/silicon and 
a thin pure Si shell, respectively. It is also clear that the early-time
evolution of the  blueshift of the line-profile minimum will be largely 
determined by the evolution of  relative strength of the two components, 
and therefore will be very difficult to interpret.

\citet{seppo_01el} suggest that the flat-bottomed 
shape of \ion{Si}{ii}\,$\lambda$6355 in \object{SN\,2001el} and its disappearance over 
a few days can be explained by the effects of scattering within a
thin region moving at the  continuum photospheric velocity, thus requiring no
absorbing HV material to produce the line shape.  
\citet{05cg} argue that 
the triangular shape of the profile in \object{SN\,2005cg}
 with an extended {\it blue} wing (see, Fig.\,\ref{f:evo})
may be due to absorption by Si in the HV
part of the ejecta. The line profile may be reproduced if the Si abundance 
slowly decreases toward high velocities, which is typical for the
delayed-detonation models  \citep{dd}.
However, these both suggestions may have difficulties to explain asymmetric line profiles 
with a stronger blue component
as observed in \object{SN\,1999ee}. \object{SN\,1999ee} is not unique.
\object{SN\,2005cf}, observed by the ESC with daily sampling 
starting from 12 days before maximum \citep{05cf},  
shows \ion{Si}{ii}\,$\lambda$6355 line that
consists of two distinct components with profile evolution similar to 
\object{SN\,1999ee}. It is therefore likely that the "peculiar"  profiles in 
\object{SN\,2001el}, \object{SN\,1990N} and \object{SN\,2005cg} are just snapshots
of this common evolutionary pattern.  In addition, 
if more SNe\,Ia like \object{SN\,2005cf} and \object{SN\,1999ee} are found,
the suggestion of \citet{05cg} that 
the SNe\,Ia with a flat-bottomed \ion{Si}{ii}\,$\lambda$6355 line may constitute 
a separate sub-class of SNe\,Ia, possibly produced from different 
progenitors and/or explosion models can be ruled out.

In the $-11$ day spectrum of \object{SN\,2003du} the \ion{Ca}{ii} H\&K  line is a
broad, single absorption  with high 
velocity of $\sim-21000$\,km\,s$^{-1}$. In the $-7.8$ days spectrum
 another, less blueshifted \ion{Ca}{ii} H\&K component is also 
visible at velocity of $\sim-10000$\,km\,s$^{-1}$.
In the subsequent spectra, the HV component decreases in strength, 
while the  low-velocity one
grows stronger.
Qualitatively the same evolution of also observed in  \object{SN\,1994D}.
In the near maximum spectra, the HV component 
is much weaker, if present at all, in \object{SN\,1994D} and \object{SN\,1998bu} than in 
\object{SN\,2003du} and \object{SN\,2001el}. It can be seen in Fig\,\ref{f:evo} that 
the strength of the HV component in \object{SN\,1994D} decreases faster than in 
\object{SN\,2003du} and \object{SN\,2001el}, thus 
qualitatively following the evolution of 
the \ion{Si}{ii}\,$\lambda$6355 HV component. On the other hand, \object{SN\,1998bu}
either lacked HV components altogether, or they disappeared faster than in \object{SN\,1994D}.
The evolution of the \ion{Ca}{ii}~IR3 line is shown for few epochs only, 
but it is evident  that a strong HV component with velocity of $\sim-21000$\,km\,s$^{-1}$ is also present 
and that this component
disappears at different time,
earliest in \object{SN\,1994D}, followed by \object{SN\,2003du}, 
and latest in \object{SN\,2001el}.
It is also interesting to note that there is 
a  segregation of SNe Ia according  to \ion{Ca}{ii}~H\&K line profile: (i)  
SNe with a single-component line at  all epochs, 
\object{SN\,2004S} \citep{kri_04s}, \object{SN\,1999ee} and \object{SN\,2002bo} 
being examples, and  (ii) SNe like \object{SN\,2003du} and  
\object{SN\,1994D} with double-component line after maximum.
In  \object{SN\,1994D} the blue component of the 
post-maximum \ion{Ca}{ii} H\&K-split is already visible in the
$-9$ spectrum as a weak feature superimposed on the broad HV
component, while in  \object{SN\,2003du} it becomes  apparent only around maximum, possibly
because the HV component remains visible longer than in \object{SN\,1994D}.
Possible identification for this feature is
 \ion{Si}{ii}\,$\lambda$3850  \citep{nugent97,lentz},
which is also supported by the identification of strong 
\ion{Si}{ii}\,$\lambda$3850  line in the early spectrum of \object{SN\,2004dt} 
\citep{wang04dt}.

Due to severe line blending it is difficult to quantify
the strength of the HV components at different epochs. However, the qualitative 
comparison strongly suggests that  
the strength of the HV components in 
the \ion{Si}{ii}\,$\lambda$6355, \ion{Ca}{ii} H\&K and \ion{Ca}{ii}~IR3 line in given SN 
are correlated and evolve similarly. 
 The HV features in the \ion{Ca}{ii} lines are stronger 
and more separated from the  lower-velocity components than in the \ion{Si}{ii}\,$\lambda$6355 line.
Comprehensive spectral modeling of the line profiles evolution is therefore needed 
to verify the two-component hypothesis for \ion{Si}{ii}\,$\lambda$6355 and further 
investigate the HV features \citep[e.g.][]{mazzali99ee}.
Such an analysis of the \object{SN\,2003du} spectra 
will be  presented elsewhere.  
Currently, there is no consensus on the origin of the HV features.
Interaction of the ejecta with circumstellar matter close to the SN \citep[e.g.][]{ger_du} or the clumpy ejecta structure found in 
some explosion models \citep[e.g.,][]{mazzali99ee,plewa,kaple} could cause the observed HV features. 
The continuum polarization 
in SNe\,Ia is typically low, but much 
higher  polarization across  the lines including the HV features is 
often observed, which favors the clumpy ejecta model rather than a global asymmetry
\citep{wang01el,wang04dt,wang_sci,leo_du}.
The HV features may thus carry information about the 3D structure of 
the ejecta and the environment
close to the SN explosion site.  Modeling of time
sequences of flux and polarization spectra \citep[e.g.,][]{kas_01el,th_00cx,wang_sci}
 may allow us to recover this 
information and help to impose additional
constraints on the SN\,Ia explosion and progenitor models.

\section{Summary}

We present an extensive set of optical and NIR observations of the 
bright nearby Type Ia \object{SN\,2003du}. The observations started 13 days
before $B$-band maximum light, and continued for 480 days after with
exceptionally good sampling.
The optical photometry was performed after the background contamination 
from the host galaxy had been removed by subtraction of template images. 
The photometry was obtained using a number of instruments with different 
filter responses. In order to properly account for deviations from 
the standard system responses, the optical photometry was calibrated by 
applying S-corrections.

Our observations show that the spectral and photometric evolution of \object{SN\,2003du} 
in both, optical and NIR wavelengths,  closely follow that of the normal SNe\,Ia. 
The luminosity decline rate parameter $\Delta m_{15}$ is found to be $1.02\,\pm0.05$, 
 the ratio between the depth of the \ion{Si}{ii}\,$\lambda$5972 and $\lambda$6355 lines 
$\mathcal{R}($\ion{Si}{ii}$)=0.22\,\pm0.02$ and the  velocity of the \ion{Si}{ii}\,$\lambda$6355 
 line is
$\sim-10000$\,km\,s$^{-1}$ around maximum light. The analysis of the $uvoir$ light 
curve suggests that $\sim0.6-0.8\,M_\odot$ of \element[][56]{Ni} was synthesized 
during the explosion. All this indicates an average normal SN\,Ia. 
We also find that \object{SN\,2003du} was 
unreddened in its host galaxy.
This property is important for better understanding of the intrinsic colors of SNe\,Ia 
in order to obtain accurate estimates of the dust extinction to the high-redshift SNe\,Ia,
which is 
one of the major 
systematic uncertainties in their cosmological use.
\object{SN\,2003du} also showed strong high-velocity features 
in \ion{Ca}{ii} H\&K and \ion{Ca}{ii}~IR3 lines, and possibly in \ion{Si}{ii}\,$\lambda$6355. 
The excellent 
temporal coverage allowed us to compare the time evolution of the line profiles with other 
well-observed SNe\,Ia  and we found evidence that the peculiar pre-maximum 
evolution of \ion{Si}{ii}\,$\lambda$6355 line in many SNe\,Ia is due to the presence of two
blended absorption components.

The well-sampled and carefully calibrated data set we present
is a significant addition to the well-observed SNe\,Ia  and the 
data will be made publicly available for further analysis. 
For example comprehensive modeling of the extensive spectral data set, 
e.g. by the abundance tomography method 
\citep{ab_tom}, may eventually help to achieve a better 
understanding of the physics of SNe\,Ia explosions and their progenitors.

\begin{acknowledgements}
This work is partly supported by the European Community's Human
Potential Program ``The Physics of Type Ia Supernovae'', under
contract HPRN-CT-2002-00303.  V.S. and A.G. would like to thank
the G\"oran Gustafsson Foundation for financial support. The work of
D.Yu.T. and N.N.P. was partly supported by the grant RFBR
05-02-17480. The work of S.M. was supported by a EURYI scheme award.

This work is based on observations collected at the Italian Telescopio
Nazionale Galileo (TNG), Isaac Newton (INT) and William Herschel (WHT)
Telescopes, and Nordic Optical Telescope (NOT), all located at the
Spanish Observatorio del Roque de los Muchachos of the Instituto de
Astrofisica de Canarias (La Palma, Spain), the 1.82m and 1.22m
telescopes at Asiago (Italy), the 2.2m and 3.5m telescopes at Calar
Alto (Spain), the United Kingdom Infrared Telescope (UKIRT) at Hawaii
and the 60-cm telescope of the Beijing Astronomical Observatory
(China).
We thank the support astronomers of these telescopes for performing
part of the observations.  We also thank the director of the Calar
Alto Observatory Roland Gredel for allocating additional
time at the 2.2m telescope in May 2003.

We thank all observers that gave up part of their time to observe 
SN~2003du.  Thomas Augusteijn and Amanda Djupvik are acknowledged for
observing during two technical nights at the NOT. Observations were
also obtained at the NOT during a student training course in
Observational Astronomy provided by Stockholm Observatory and the
NorFA Summer School in Observational Astronomy. We thank Geir Oye for
excellent support and close collaboration during this course. We also
thank O.A.Burkhanov, S.Yu.Shugarov and I.M.Volkov for carrying out
observations at Maidanak, Slovakia, Moscow and Crimea. We thank Cecilia Kozma for
making available to us her late-time light curve models. We thank Aaron
Barth for providing us with the spectra of SN 1994D collected by the
Alexei Filippenko group at UC Berkeley, and the people who did the
observations: Aaron Barth, Alexei Filippenko, Tomas Matheson, Xiaoming
Fan, Michael Gregg, Vesa Junkkarinen, Brian Espey, Matt Lehnert, Lee
Armus, Graeme Smith, Greg Wirth, David Koo, Abe Oren and Vince
Virgilio. We thank K. Motohara and the co-authors of  
\citet{03du_late_ir} for providing us with the unpublished late-time
NIR magnitudes of SN~2003du.

This work has made use of the NASA/IPAC Extragalactic Database (NED),
the Lyon-Meudon Extragalactic Database (LEDA), 
NASA's Astrophysics Data System, the SIMBAD database operated at CDS, Strasbourg, France,
data products from the Two Micron All Sky Survey and the SUSPECT supernova spectral archive.

\end{acknowledgements}

%\bibliographystyle{aa}   % if natbib is available
%\bibliographystyle{aipprocl} % if natbib is missing

%%%%%%%%%%%%%%%%%%%%%%%%%%%%%%%%%%%%%%%%%%%
%% You probably want to use your own bibtex database here
%%%%%%%%%%%%%%%%%%%%%%%%%%%%%%%%%%%%%%%%%%%
%\bibliography{/data2/sn2003du/paper/AA/sn}

\appendix

\section{S-corrections}

\subsection{Photometric systems responses}

The most important ingredient for computing S-corrections is the accurate 
knowledge of the object SED and the response of the instruments used. 
\citet{ph02er} have calculated the responses of most of 
the instruments used by the ESC.  
However, we repeated the process for the 5
instruments most frequently used to observe \object{SN\,2003du}: AFOSC
at Asiago 1.82m telescope, DOLORES at TNG, ALFOSC at NOT, CAFOS at
Calar Alto 2.2m telescope and BAO 60-cm telescope imager, using the
new extensive spectrophotometry of Landolt stars in \citet{specphot}. The first four of the five instruments 
are combined spectrographs/imagers with
design that allows the 
grisms, the slits and the imaging filters to be inserted into the light beam 
simultaneously. 
This made possible to measure the filter transmissions {\it in situ} as the 
filters are mounted in the instrument and used during the photometric 
observations.  This measurement is straightforward and
consists of taking spectral flat-fields with and without the filter in
the beam. The flat taken with the filter is divided by the one taken
without, giving the filter transmission.  Before doing this, the bias
and any reflected light present was carefully removed.  The latter can
be important in the blue part of the spectrum where the sensitivity of
the system is low and the scattered light can be a significant
fraction of the useful signal; this can affect the measured
transmission.  The wavelength calibration was done with arc-lamp
spectra taken without the filter in the beam. When filters are
introduced in the beam small shifts of the wavelength solution can be
expected. After the measurements we checked this for few
filters at AFOSC at Asiago 1.8m telescope and did indeed find shifts
of a few pixels. Hence, the measured filter transmissions might be
shifted by up to 20--30\,\AA, but the shape is accurately determined.
Generally, we found good agreement with the filter transmissions
available from the instrument web-pages.  However, we found
significant discrepancies for the Calar Alto 2.2m + CAFOS $B$ and $I$
filters, and minor differences for the TNG+DOLORS $I$-band.  For the
$U$-bands we used the transmissions available from the instrument
web-pages. For the BAO 60-cm telescope the filter transmissions
specified by the manufacturer were used.

 \begin{figure*}[!th]
 \centering
 \includegraphics*[width=16cm]{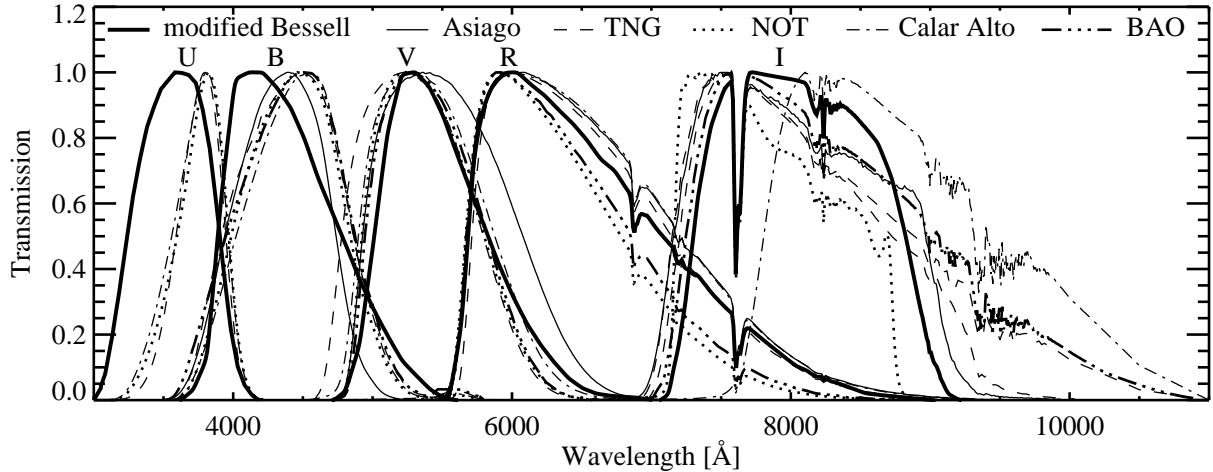}
 \caption{Reconstructed system responses for the five instruments studied 
compared to the  modified Bessell passbands. }
 \label{f:filters}
 \end{figure*}

The total system responses were computed by multiplying the filter
transmissions by (a) the CCD quantum efficiency (QE), (b) the
reflectivity of at least two aluminum surfaces, (c) the continuum
transmission of the Earth atmosphere at airmass one (the extinction
laws were provided by the observatories), and (d) a telluric
absorption spectrum, which we derived from the spectrophotometric 
standards observed at WHT close to airmass one. The lens and window transmissions
were not included because this information was unavailable. 
Synthetic magnitudes were calculated from \citet{specphot}
spectrophotometry of Landolt standard stars,
$m^{syn}=-2.5\log\left(\int\,f_\lambda^{phot}(\lambda)R^{nat}(\lambda)d\lambda\right)$.
The difference between the synthetic and the observed photometry was fitted as a function of the 
observed color indices to compute synthetic color-terms ($ct^{syn}$), e.g. for $B$ we have
\begin{equation}
B^{std}-B^{syn}=ct^{syn}(B^{std}-V^{std})+const.
\label{eq:synthcol}
\end{equation}
For the $VRI$-bands, the $ct^{syn}$'s were close to the observed
ones $ct^{obs}$. In some cases small differences exceeding the uncertainty were
accounted for by shifting the filter transmissions until $ct^{syn}$
matched $ct^{osb}$.  Small shifts of up to $\sim20-30$\,\AA\ were
required. These discrepancies could easily have arisen from the way in
which the transmissions were determined, as discussed above. For the
$U$ and $B$-bands however, we found large differences which would have
required an unacceptably large shift to correct for them.  The
synthetic $U$ and $B$ bands were always too blue. This is, to some
extent, to be expected because the neglected optical elements like lenses
or windows, anti-reflection and other coatings will tend to reduce the
system sensitivity shortward of $\sim$4000\,\AA.  The uncertainty in
the CCD QEs and the extinction laws may also contribute to this
effect. To account for the net effect of these uncertainties we
modified the $U$ and $B$ bands by multiplying them with a smooth
monotonic function of wavelength so that $ct^{syn}$ matched
$ct^{obs}$. We used the Sigmoid function
\begin{equation}
F(\lambda;\lambda_0,\Delta)=\frac{1}{1+\exp(-(\lambda-\lambda_0)/\Delta)},
\label{eq:sig}
\end{equation}
that changes smoothly from 0 to 1. The parameters $\lambda_0$ and
$\Delta$ control the position and the width of the transition; for
small $\Delta$ the Sigmoid function approaches a Heaviside step
function at $\lambda_0$.  We proceeded as follows: $\lambda_0$ and
$\Delta$ were varied in the wavelength intervals 3200--4200\,\AA\ and
100--500\,\AA, respectively, and the set of parameters that brought
the synthetic $U$ {\it and} $B$-band color-terms into accord with the
observed ones was chosen. Note that independent modification of $U$
and $B$ results in degeneracy in the ($\lambda_0$,$\Delta$) parameter space,
and it was only when the $U$ and
$B$-bands were considered together that an unique solution for $\lambda_0$ and
$\Delta$ could be obtained.  As standard Johnson-Cousins system responses we use 
the \citet{ubvri} filters but following \citet{specphot} we first modified them so that they could be
used with photon fluxes and included the telluric absorptions. Small
shifts were also applied to account for the small color-terms that
are noticeable when compared with the Landolt photometry.
\citet{besscorr} suggested correcting Landolt
photometry to bring it into the original Cousins system. The synthetic
photometry with the original Bessell filters does match the corrected
magnitudes. However, for sake of comparability with the existing SN
photometry, we use the original Landolt photometry and modify the
Bessell filters so that the synthetic color-terms are zero.  The
constant terms derived from the fits with Eq.\,\ref{eq:synthcol} are the
filter zero-points for synthetic photometry.  The constant in
Eq.\,\ref{eq:scorr} is the difference between the zero-point for the
Bessell and natural passbands.

The reconstructed bands are shown in Fig.\,\ref{f:filters}  together with
the modified Bessell filters, demonstrating  the variety of passbands one 
may encounter at different telescopes. Note particularly the non-standard 
form of the Calar Alto $I$-band and NOT $R$-band.
We note that the reconstructed responses should be regarded only as
approximations of the real responses. A given passband can be modified in
many ways to match the observed and the synthetic color-term, and we
would consider the procedure we used as the most appropriate one
given the available information.
We also note that fitting the $U$-band synthetic
color-term is ambiguous.  Because of the Balmer discontinuity
even small deviation from the Bessell passband changes
$U^{std}-U^{syn}$ such that it needs no longer be a simple linear
function of $U^{std}-B^{std}$.  This also affects the 
derivation of the observed color-terms and 
as a
result the $U$-band photometry should be in general considered significantly less
accurate than other bands.

%%%%%%%%%%%%%%%%%%%%%%%%%%5

\subsection{Computing the S-corrections}

We used our spectra obtained earlier than 110 days after maximum, and spectra from 
\citet{ger_du} and \citet{anu_du} to compute the 
S-corrections and to transform  the
$BVRI$ photometry of \object{SN\,2003du} into the
Johnson-Cousins system. The TNG, Calar Alto and BAO $I$-bands 
 extend out to 1.1\,$\mu$m and to compute the  S-corrections,  
we also used our NIR spectra of \object{SN\,2003du} (Sec.\,2.3).
 To compute the BAO $I$-band S-corrections between  $+30$ and $+63$ days we
also used NIR spectra of \object{SN\,1999ee} \citep{mario99ee} and
\object{SN\,2000ca} (Stanishev et al., in preparation) taken at $\sim+40$ days. 
 The $U$-band could not be
S-corrected because no UV spectra of \object{SN\,2003du} were
available.

The relative spectrophotometry of
\object{SN\,2003du} was not always sufficiently accurate for the purpose of
computing S-corrections. It was thus necessary to
slightly modify some of the spectra so that the synthetic photometry with the 
modified Bessell $BVRI$ bands matched the observed one. 
To achieve that, the spectra were multiplied by a smooth correction function 
determined by fitting the ratio between the observed and
the synthetic fluxes.
When the ratio varied monotonically with wavelength, a second-order
 polynomial was used.  When a more complex function was required, 
a spline fit was used.
At the first iteration the
synthetic magnitudes were compared with the linear color-term corrected  
magnitudes of \object{SN\,2003du},  
and the spectra were only modified if the observed and the
synthetic color indices differed by more than 0.05 mag for $B-V$ and
$V-R$, and 0.1 mag for $V-I$. These corrected spectra were used to
compute S-corrected photometry of \object{SN\,2003du}.
The flux correction of the spectra was then repeated using the S-corrected 
rather than the color-term corrected
photometry.  Spectra were only
corrected if the color discrepancies were greater than 0.03 mag
for $B-V$ and $V-R$, or 0.05 mag for $V-I$. New S-corrected photometry was
then computed and the process repeated
to obtain the final S-corrected photometry and calibrated spectra. A
number of spectra have a wavelength coverage that only allows $B$ and
$V$ synthetic magnitudes to be computed.  In these cases, 
only a simple linear correction was applied to match the observed $B$
and $V$ magnitudes.

We note that because the instrumental responses are fairly close to
those of Bessell filters, the S-corrections are almost entirely
determined by the SN spectral features and are practically insensitive
to small changes of the SN colors.  It was found that the
initial correction of the spectra yielded spectrophotometry which was
already accurate to a few per cent and that the subsequent iterations had
very little effect on the final calibrated photometry.  We therefore
conclude that the few percent uncertainties in the spectrophotometry,
which might have arisen from the way the spectra were corrected,
should have little effect on the final photometry.

\end{document}